\documentclass[apj,iop]{emulateapj}

\usepackage[table,xcdraw]{xcolor}
\usepackage{diagbox}
\usepackage{amssymb}
\usepackage{mathtools}
\usepackage{color}
\usepackage{xifthen}
\newcommand{\toto}[1][]{ %
\ifthenelse{\isempty{#1}}{{\color{red} Missing}}{{\color{red} #1}} %
}

\usepackage[]{bm}

\shorttitle{ACAD-OSM II}
\shortauthors{J. Mazoyer et al.}
\usepackage[colorlinks,breaklinks=false, citecolor = {blue}, urlcolor = {blue}]{hyperref}

\begin{document}

\title{Active correction of aperture discontinuities - optimized stroke minimization II: optimization for future missions}

\author{J.~Mazoyer\altaffilmark{1,2}, L.~Pueyo\altaffilmark{2}, M.~N'Diaye\altaffilmark{3}, K.~Fogarty\altaffilmark{1}, N.~Zimmerman\altaffilmark{2}, R.~Soummer\altaffilmark{2}, S.~Shaklan\altaffilmark{4} and C.Norman\altaffilmark{1}}

\altaffiltext{1}{Johns Hopkins University, Zanvyl Krieger School of Arts and Sciences, Department of Physics and Astronomy, Bloomberg Center for Physics and Astronomy, 3400 North Charles Street, Baltimore, MD 21218, USA}
\altaffiltext{2}{Space Telescope Science Institute, 3700 San Martin Dr, Baltimore MD 21218, USA}
\altaffiltext{3}{Universit\'e C\^ote d\'{}Azur, Observatoire de la C\^ote d\'{}Azur, CNRS, Laboratoire Lagrange, Bd de l\'{}Observatoire, CS 34229, 06304 Nice cedex 4, France}
\altaffiltext{4}{Jet Propulsion Laboratory, 4800 Oak Grove Drive, Pasadena, CA 91109, USA}

\email{jmazoyer@jhu.edu}

\begin{abstract}

{High-contrast imaging and spectroscopy provide unique constraints for exoplanet formation models as well as for planetary atmosphere models. Instrumentation techniques in this field have greatly improved over the last two decades, with the development of stellar coronagraphy, in parallel with specific methods of wavefront sensing and control. Next generation space- and ground-based telescopes will allow the characterization cold solar-system like planets for the first time and maybe even \textit{in situ} detection of bio-markers.}
{However, the growth of primary mirror diameters, necessary for these detection, comes with an increase of their complexity (segmentation, secondary mirror features). These discontinuities in the aperture can greatly limit the performance of coronagraphic instruments. In this context, we introduced a new technique, Active Correction of Aperture Discontinuities - Optimized Stroke Minimization (ACAD-OSM), to correct for the diffractive effects of aperture discontinuities in the final image plane of a coronagraph, using deformable mirrors.}
{In this paper, we present several tools that can be used to optimize the performance of this technique for its application to future large missions.}
{In particular, we analyze the influence of the deformable setup (size and separating distance) and found that there is an optimal point for this setup, optimizing the performance of the instrument in contrast and throughput while minimizing the strokes applied to the deformable mirrors.}
{These results will help us design future coronagraphic instruments to obtain the best performance.}
\end{abstract}

\keywords{instrumentation - coronagraphy - exoplanets - high-contrast - direct imaging}

%-------------------------------------------------------------------------------------------------
%-------------------------------------------------------------------------------------------------
%-------------------------------------------------------------------------------------------------
\section{Introduction}
\label{sec:intro}
%-------------------------------------------------------------------------------------------------
%-------------------------------------------------------------------------------------------------
%-------------------------------------------------------------------------------------------------

\subsection{Context}

The current generation of high-contrast imaging instruments \citep{macintosh08, beuzit08, martinache10,oppenheimer12} have already detected several exoplanets \citep{marois08, lagrange09, Kuzuhara13, rameau13b, bailey14, macintosh15} and obtained their first spectra \citep{bonnefoy16,chilcote17,rajan17}. Installed on ground-based 8m class telescopes, these facilities benefit from important improvements in adaptive optics \citep[AO, ][]{sauvage07, wallace09} and coronagraphy \citep{soummer03,rouan00} to reach $10^{-6}$ contrast levels and image young, bright Jovian planets. The characterization of these nearby exoplanets has already helped us to better understand the formation and evolution of exo-planetary systems. 

Coronagraphs on these current generation of ground-based Extreme-AO systems were designed taking into account the central obscuration, but not the secondary support structures \citep{soummer05}. With the emergence of future projects relying on telescope geometries that can be heavily segmented, or with large secondary and spiders either from the ground \citep[Extremely Large Telescopes, ELTs, ][]{macintosh06, kasper08,davies10,quanz15c} or from space, such as LUVOIR  \citep[Large UV/Optical/IR Surveyor,][]{dalcanton15} and WFIRST \citep[Wide Field Infrared Survey Telescope,][]{spergel15}, the need for solutions that optimize coronagraph performances for all features in the aperture is becoming more pressing. 

Several static (e.g relying on printed phase/amplitude masks in the pupil/image plane of the instrument) coronagraph solutions have been proposed over the past few years: APLC \citep[Apodized pupil Lyot coronagraph,][]{soummer07,ndiaye16}, PIAACMC \citep[Phase-Induced Amplitude Apodization complex mask coronagraph,][]{guyon2005, Guyon14}, or apodized Vortex \citep{ruane16,fogarty17}. We chose another approach and introduced a new technique, Active Correction of Aperture Discontinuities - Optimized Stroke Minimization (ACAD-OSM), in a previous paper \cite{mazoyer17a} --hereafter ACAD-OSM I. This method relies on deformable mirrors (DMs), to compensate for the diffraction associated with spiders and segment gaps in a high-contrast imaging instrument. This DM-based solution was inspired by the ACAD-ROS (Active Compensation of Aperture Discontinuities - Ray Optics Solution) technique presented in \cite{pueyo13}, but uses a different algorithm to calculate the deformation of the active surfaces. ACAD-ROS was using a geometric optics approach to ``fill the gaps" in the pupil. ACAD-OSM, on the other hand, directly aims at correcting for the diffractive effects of aperture discontinuities to create a dark hole (DH) in the final image plane of a coronagraph. It is derived from a linear correction algorithm \citep[stroke minimization, ][]{pueyo09}, designed to correct for small phase and amplitude aberrations and using an interaction matrix linking DM commands to their effects in the science focal plane. However, because of the important strokes necessary to correct for the effects of aperture discontinuities, this kind of linear algorithms quickly diverge in this case. ACAD-OSM is therefore re-computing the interaction matrix several times as the correction process changes the wavefront shape. This approach makes it very similar to linear active algorithms that were developed to correct for ``small'' phase and amplitude aberrations \citep[e.g.][]{baudoz06,borde06,giveon07,riggs16}, except that the DMs can now reach strokes of up to several hundred nanometers, which are necessary to correct for aperture discontinuities. In ACAD-OSM I, we presented the essence of this new algorithm and highlighted its key features:

\begin{itemize}
    \item The contrast achievable with ACAD-OSM can be orders of magnitude deeper than using ACAD-ROS for a variety of telescope apertures. The capabilities of this technique are illustrated by showing that for three aperture geometries (WFIRST, E-ELT and LUVOIR-like) the final contrast for a 10\% bandwidth (BW) was at least $10^{-9}$. 
    \item The contrast obtained by the ACAD-OSM solution is a weak function of the BW, suggesting that the ultimate BW of such and instrument will be driven by the correction of wavefront errors, not aperture discontinuities. More importantly, we identified that there exists a well behaved regime of DM setups for which the chromatic response of ACAD-OSM follows the theoretical predictions from \citet{shaklan06}. 
    \item  Wavefront control commands can be superposed on to DM shapes calculated with ACAD-OSM, thus providing a simple active framework to compensate simultaneously for aperture discontinuities, surface figure errors, small segment and coronagraph elements misalignments as well as for phase and amplitude errors.       
\end{itemize}

However, in ACAD-OSM I, we did leave open three important questions. First, we did not explore the trades between contrast (e.g how much does the stellar host is suppressed by the instrument) and throughput (e.g how much of the planet flux is propagated through the instrument). Such trades are extremely relevant when trying to optimize the Signal to Noise Ratio (SNR) of an exo-planet and depend both on the choice coronagraph architecture and of the ACAD-OSM DM shapes. Second, we only suggested that the geometry of the DM setup (size of the DMs, separation between them) had an influence on the final performance of ACAD-OSM, by showing a few examples of possible setups. In reality, the study of the whole continuum of solutions, which has been recently presented in the case of ``small'' wavefront aberration control by \citet{beaulieu17}, is primordial to push this technique to its limits. It remains to be studied in the more challenging case of aperture discontinuities. Third, we did not discuss at all the impact of low order spatial aberrations, namely jitter and stellar angular size on the final contrast after ACAD-OSM. In the present paper, we tackle these questions in order to provide a comprehensive portrait of the advantages and limitations of the ACAD-OSM technique in the context of future large missions.

After briefly summarizing in Section~\ref{sec:metrics} the metrics that will be used to optimize the performance of this technique, we will start by comparing the performance obtained by ACAD-OSM with different coronagraphs on the same aperture in Section~\ref{sec:scda}. We will then describe in Section~\ref{sec:fine_tunning}, for a given aperture and DM setup, how the contrast or the throughput performance of the resulting DH can be preferentially optimized to enhance the SNR of the system. We will then address the most important point of this article, the influence of the DM setup on the performance of ACAD-like techniques, first theoretically  in Section~\ref{sec:formalism} and then by exploring the parameter space using numerical simulations in Section~\ref{sec:DM_setup}. Finally, in Section~\ref{sec:optim_missions}, the tools presented in this paper will be used to provide an example of how to optimize the ACAD-OSM technique in the case of the WFIRST mission. This paper comes with a long appendix, describing analytically the fundamental limitations of 2 DM correction techniques.

%-------------------------------------------------------------------------------------------------
\begin{figure}
\begin{center}
 \includegraphics[width = .48\textwidth]{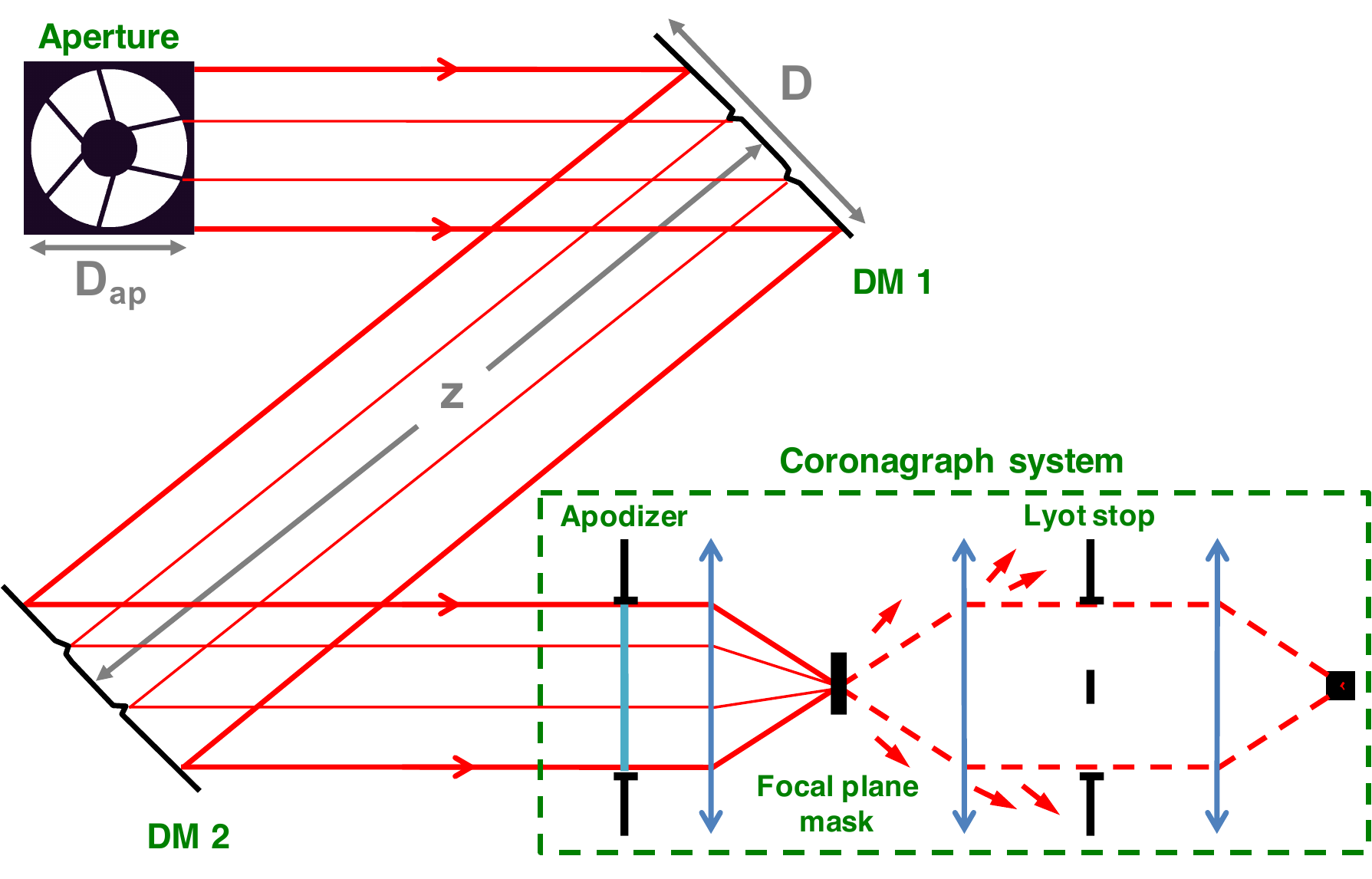}
 \end{center}
\caption[plop]
{\label{fig:schema_ACAD} Schematic representation of a two DM system and a coronagraph. The distances z, D and $D_{ap}$ are shown on this optical layout.}
\end{figure}
%-------------------------------------------------------------------------------------------------

%-------------------------------------------------------------------------------------------------
\subsection{Description of performance metrics}%the simulations and performance metrics}
\label{sec:metrics}
%-------------------------------------------------------------------------------------------------

All the results presented in this article have been obtained following the same method. First, the corrective DM shapes are calculated using the algorithm described in ACAD-OSM I. The wavefront estimation in focal plane is assumed to be perfect. Several method have been developed to measure the electrical field produced by both amplitude and phase errors directly in the focal plane \citep{mazoyer14,riggs16}. For large BWs, the interaction matrix is the concatenation of several matrices built at different discrete wavelengths equally sampled over the BW. The chosen number of discrete wavelengths depends on the BW (see ACAD-OSM I). For a 10\% bandwidth, 3 discrete wavelengths are used. Once the DM shape solutions are obtained, a larger number of wavelengths is used, centered around the central wavelength and equally spatially sampled one to another over the BW (usually more than 1 per percent of the BW) to create the final focal plane image on which the performance are measured. In this paper, several metrics are used to describe the performance of the ACAD-OSM technique.

\begin{itemize}
    \item \textit{Contrast}. The contrast is defined as the residual star intensity at a given point of the final focal plane, normalized by the maximum of the point spread function (PSF) obtained on-axis in the final focal plane when the coronagraph focal plane mask (FPM) is removed. We frequently show contrast curves, \textit{i.e.} radial profiles of the azimuthal mean contrast in the focal plane as a function of the distance to the star (in telescope resolution elements $\lambda_0/D_{ap}$) or just print the mean contrast in the DH. 
    
    \item \textit{Transmission and throughput}. Several metrics are used to measure the impact of the design on off-axis source detection. The transmission of a system is the percentage of the light energy not blocked during the propagation through this system. In this article, we separate the coronagraph transmission (whose optimization is out of the scope of this paper), from the 2 DM system transmission. However, to take into account the complex effects of the ACAD-OSM technique on an off-axis PSF, we also define the off-axis throughput: for a given separation of an off-axis source, what percentage of the total light hitting the primary mirror of the telescope is finally in the core of the PSF in the final focal plane (in a radius of 0.7 $\lambda_0/D_{ap}$ around the expected position of the PSF).
    
    \item \textit{Robustness to low order aberration residuals}. The performance of a space- or ground-based telescope is often limited by the stability of the star PSF during the observation. When the star PSF moves outside of the optical axis, the performance of the coronagraph quickly degrades. In this paper, only the robustness of a coronagraph to small tip-tilt (TT) jitter in $\lambda_0/D_{ap}$ is measured.
\end{itemize}
In this paper, we will discuss performance only for the correction of aperture discontinuities, and we will not consider phase errors. A recent and complementary paper \citep{beaulieu17} studied a similar two-DM problem assuming a realistic distribution of phase and amplitude errors. Also, as in ACAD-OSM I, coronagraphs with a static apodization have been optimized to correct for the central obscuration and ACAD-OSM method is only correcting for the effects of discontinuities: secondary mirror support structures and segmentation.

Finally, note that in this paper, we choose not to directly discuss the impact of the ACAD-OSM on the Inner and Outer Working Angles (IWA, OWA). To the first order these performance metrics are related to the static component of the coronagraph setup as follows:
\begin{itemize}
    \item The outer radius of the DH ($OWA_{DH}$) is bounded by the number of actuators across the beam, just as in classical wavefront control. This is true unless the static coronagraph masks accommodating for the central obscuration have an $OWA_{DH}$ smaller than $N_{act}/2$.
    
    \item The inner radius of the DH ($IWA_{DH}$) is bounded by the static coronagraph FPM accommodating for the central obscuration. That is, the ACAD-OSM solution can only attain a high contrast as close to the star as allowed by the static design on an azimuthally symmetric aperture. In the case of coronagraphs for which $IWA_{DH} \sim \lambda_0/D_{ap}$ (e.g. vortex type coronagraph) then we arbitrarily set $IWA_{DH}$ to a few $\lambda_0/D_{ap}$. \citet{ruane16}, as well as ACAD-OSM I, showed that for such static coronagraphs, DHs usually extended inwards to $\lambda_0/D_{ap}$ regardless.  
\end{itemize}

However, the more astrophysical definitions of these quantities ($IWA_{planet}$ and $OWA_{planet}$) are the focal plane radii between which the transmissivity of the planet signal is ``sufficient enough,'' and depend on subtle trades made in the choice of the static coronagraph masks along with ACAD-OSM DM setup and shapes. In the absence of literature consensus regarding the definition of ``sufficient enough,'' we choose to present the results of this paper using a more general format and we will usually display 1D curves relating encircled energy within $0.7 \lambda_0/D_{ap}$ of the planet to angular separation between the planet and on-axis star in $\lambda_0/D_{ap}$. For this reason the notation IWA and OWA in this paper will refer to $IWA_{DH}$ and $OWA_{DH}$, the chosen edges of the DH.

%-------------------------------------------------------------------------------------------------
\begin{table*}
\centering
\caption{Parameter of the 3 types of coronagraphs used in this article for the SCDA and WFIRST aperture.}
\label{tab:ularasa}
\begin{tabular}{|c|c|c|c|c|c|}
\hline
              & FPM              & LS inner radius &  LS outer radius & Optimized for                & Transmission\\ \hline  \hline
\multicolumn{6}{|c|}{36\% central obscuration (WFIRST)}                                                                                                                                                               \\ \hline  \hline
APLC   & Opaque Lyot mask: 5 $\lambda/D_{ap}$ radius & 50\%     & 100\%   & \begin{tabular}[c]{@{}c@{}}c = $10^{-10}$ (10\% BW)\end{tabular}  & 45\%                   \\ \hline
RAVC4         & Phase charge 4 vortex & 78\%         & 100\%             & ideal                                        & 26\%                   \\ \hline
SAVC6 & Phase charge 6 vortex    & 55\%         & 100\%              & ideal                                & 72\%                   \\ \hline  \hline
\multicolumn{6}{|c|}{17\% central obscuration (SCDA)}                                                                                                                                                                 \\ \hline  \hline
APLC          & Opaque Lyot mask: 4 $\lambda/D_{ap}$ radius & 30\%   & 92\%               & \begin{tabular}[c]{@{}c@{}}c = $10^{-10}$ (10\% BW)\end{tabular} & 58\%                   \\ \hline
RAVC4         & Phase charge 4 vortex    & 56\%          & 100\%             & ideal                                                           & 58\%                   \\ \hline
SAVC6 & Phase charge 6 vortex    & 41\%        & 100\%            & ideal                                                           & 76\%                   \\ \hline 
\end{tabular}
\end{table*}
%-------------------------------------------------------------------------------------------------

%-------------------------------------------------------------------------------------------------
%-------------------------------------------------------------------------------------------------
%-------------------------------------------------------------------------------------------------

\section{Impact of the static coronagraph choice, case of SCDA pupil}
\label{sec:scda}
%-------------------------------------------------------------------------------------------------
%-------------------------------------------------------------------------------------------------
%-------------------------------------------------------------------------------------------------

%-------------------------------------------------------------------------------------------------
\begin{figure}
\begin{center}
 \includegraphics[width = .48\textwidth, trim= 0.1cm 4.5cm 5cm 4cm, clip = true]{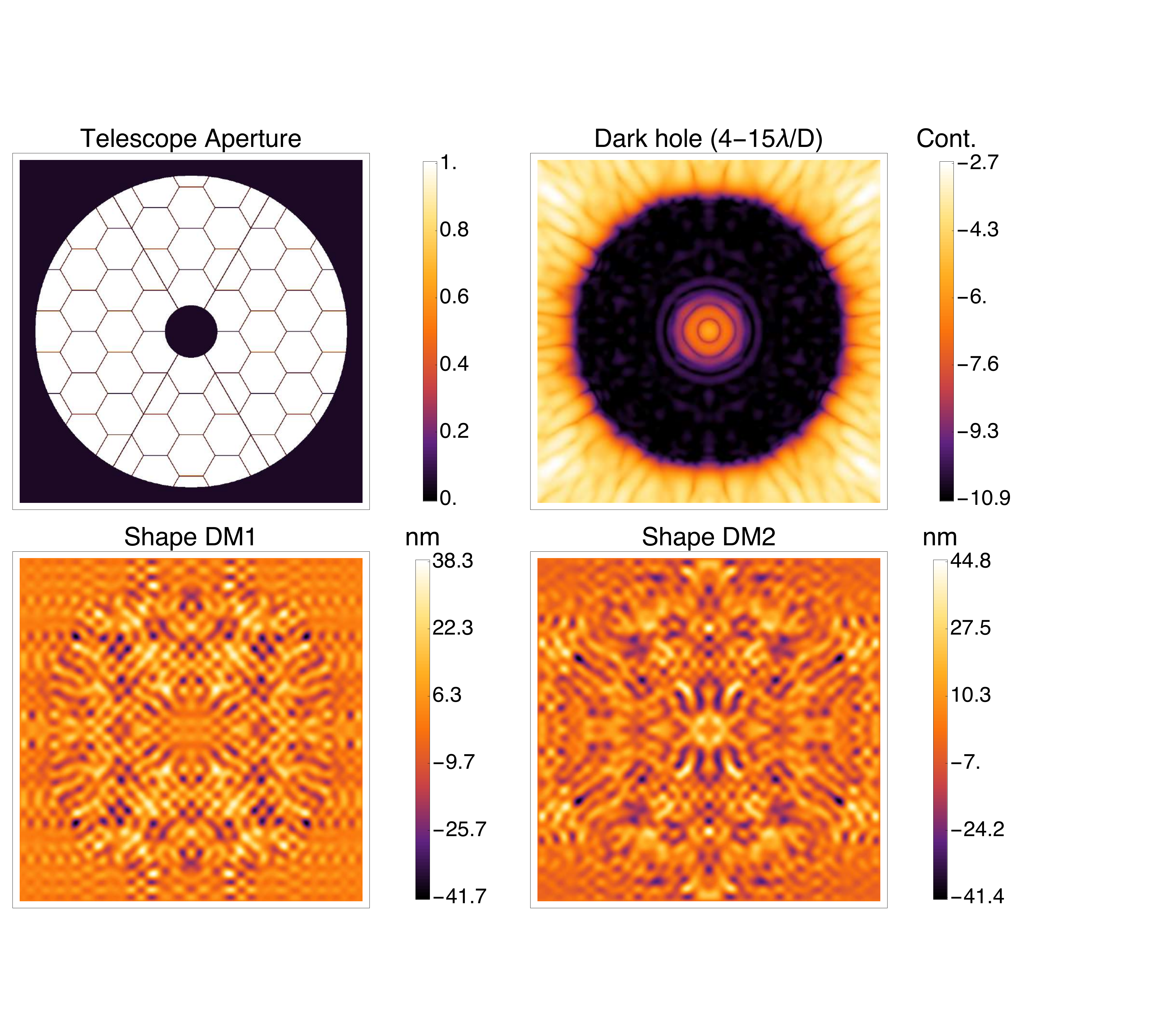}
 \end{center}
\caption[]
{\label{fig:scda_dh} Results for the SCDA aperture (APLC coronagraph, $N_{act} = 48$, IAP = 0.3 mm, D = $48 * 0.3$ mm, $z = 0.3$ m and $\Delta \lambda /\lambda_0 = $ 10\% BW). Top left: SCDA aperture. Top right: the final DH. Bottom: the DM shapes obtained using ACAD-OSM for this solution.}
\end{figure}
%-------------------------------------------------------------------------------------------------

Figure~\ref{fig:schema_ACAD} shows the schematic of the two DM system coupled with a coronagraph used by ACAD-like methods and ACAD-OSM in particular. This schematic introduces the diameter of the aperture $D_{ap}$ and the distance between the DMs $z$. The two DMs are assumed to be square with area $D\times D$, and to have $N_{act}\times N_{act}$ actuators. The size $D$ of the DM will often be written in the form $N_{act} *$ IAP, where IAP is the DM Inter-Actuator Pitch. In this paper, we assume that there is no limit on the amount of strokes achievable on the DMs.

In this section, ACAD-OSM is used in conjunction with several coronagraphic designs and compare the results. This comparison is made using an aperture from the Segmented Coronagraph Design and Analysis (SCDA) program\footnote{Program led by Stuart Shacklan (JPL). Please check: \url{https://exoplanets.nasa.gov/system/internal_resources/details/original/211_SCDAApertureDocument050416.pdf}} which aims at evaluating the performance of coronagraphic instruments with different space telescope apertures. This aperture is shown in Figure~\ref{fig:scda_dh} (top, left). 

The first coronagraph studied is an apodized pupil Lyot coronagrah APLC \citep{soummer03,soummer11, ndiaye16}. The apodization, the radius of the opaque Lyot FPM, and the Lyot stop inner and outer radius have been optimized to maximize the throughput, while providing a $10^{-10}$ contrast over a BW of 10$\%$. The second coronagraph is a charge 4 ring-apodized vortex coronagraph \cite[RAVC,][]{mawet13}. For this coronagraph, the size of the Lyot stop inner radius is measured using the analytic method developed in \cite{mawet13}, which drastically limits the throughput for the charge 4 coronagraph. For this reason, a third coronagraph is used in this study, a charge 6 polynomial apodized vortex coronagraph \citep[PAVC,][]{fogarty17}. This coronagraph also uses a vortex FPM in focal plane, but uses a different apodization. These apodizations were designed to analytically cancel the on-axis light on apertures with central obscuration, as the classical vortex coronagraph does on clear apertures. The vortex coronagraphs are simulated using the technique described in \citet[][]{mazoyer15}, in order to totally cancel the on-axis light in the Lyot stop plane at all wavelengths in the absence of wavefront aberrations or pupil discontinuities other than the central obscuration and for point-like stars. 
This assumption ensures that the effects of discontinuities in the aperture and the performance of ACAD-OSM to correct for them are the only limitations, not the numerical errors due to the simulation of the coronagraphs.

%-------------------------------------------------------------------------------------------------
\begin{figure}
\begin{center}
 \includegraphics[width = .48\textwidth]{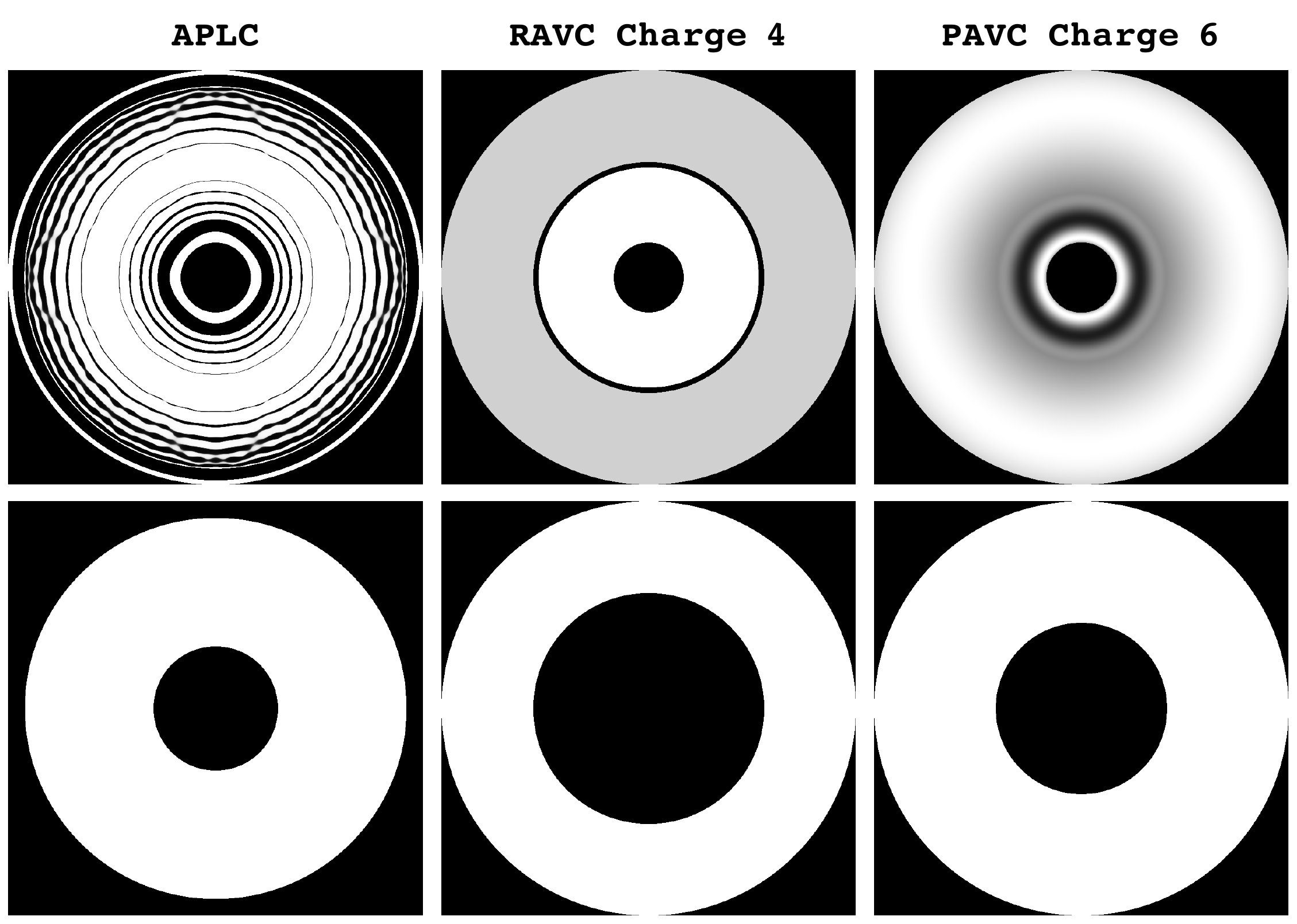}
 \end{center}
\caption[]
{\label{fig:apod_lyot_wfirst} Apodization (top) and Lyot stop (bottom) for a 17\% central obscuration (corresponding to the SCDA aperture used in this paper) for several coronagraphs. The parameters are summarized in Table~\ref{tab:ularasa}.}
\end{figure}
%-------------------------------------------------------------------------------------------------

The shape of the apodization (top) and of the Lyot stop (bottom) used in this section (for a 17\% central obscuration) are shown for these three coronagraphs in Fig.~\ref{fig:apod_lyot_wfirst} and their parameters are summarized in Table~\ref{tab:ularasa}. The IWA of the DH is set at $1 \lambda_0/D_{ap}$ for the two apodized vortex coronagraphs and $4 \lambda_0/D_{ap}$ for the APLC, because the size of the FPM is preventing us to access smaller IWAs anyway. The chosen DM setup is $N_{act} = 48$, IAP = 0.3 mm, D = $48 * 0.3$ mm, $z = 0.3$ m and $\Delta \lambda /\lambda_0 = $ 10\% BW.

\begin{figure}
\begin{center}
 \includegraphics[trim= 1.5cm 0.8cm 1.0cm 0.5cm, clip = true,width = .48\textwidth]{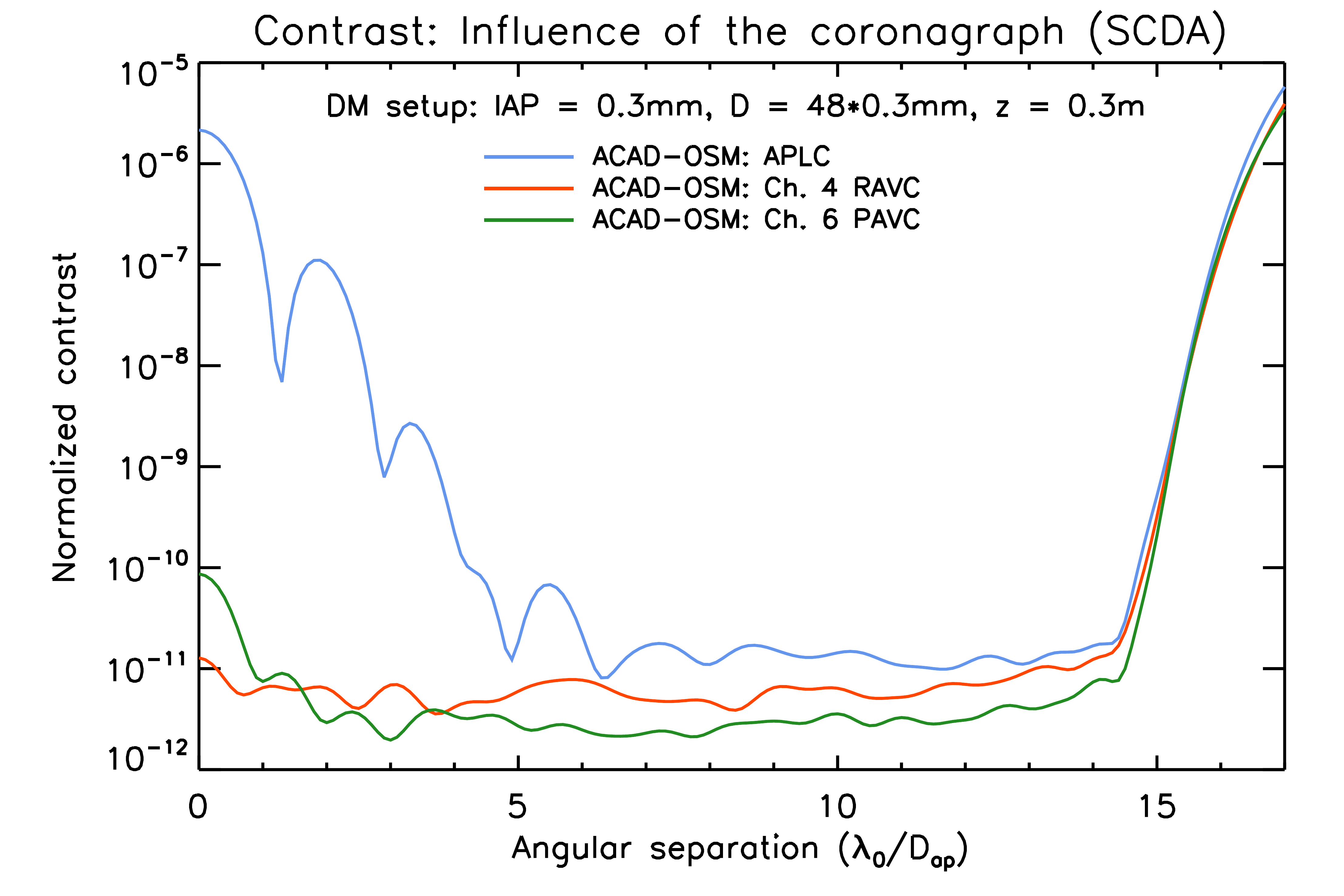}
 \end{center}
\caption[]
{\label{fig:scda_contrast} Contrast performance for the SCDA aperture ($N_{act} = 48$, IAP = 0.3 mm, D = $48 * 0.3$ mm, $z = 0.3$ m, $\Delta \lambda /\lambda_0 = $ 10\% BW), for an APLC (blue curve), a charge 4 RAVC (red curve) and a charge 6 PAVC (green curve).}
\end{figure}

Fig.~\ref{fig:scda_dh} shows the DH obtained for the SCDA aperture with an APLC coronagraph,  and the DM shapes solution of ACAD-OSM. Fig.~\ref{fig:scda_contrast} shows the results in contrast for the 3 coronagraphs. The APLC contrast, after ACAD-OSM, is limited to a level slightly better than $10^{-10}$, which is due to the fact that this coronagraph apodization has been optimized for this contrast. This show that the ACAD-OSM technique is, in that case, therefore limited by the performance of the coronagraph itself. The two apodized vortex coronagraphs were simulated to produce complete cancellation of the star and their performance contrast are very similar. We can therefore assume that this contrast level is set by the ACAD-OSM technique itself, for this aperture, DM setup and DH size. 

\begin{figure}
\begin{center}
 \includegraphics[trim= 2.5cm 0.8cm 1.0cm 0.5cm, clip = true,width = .48\textwidth]{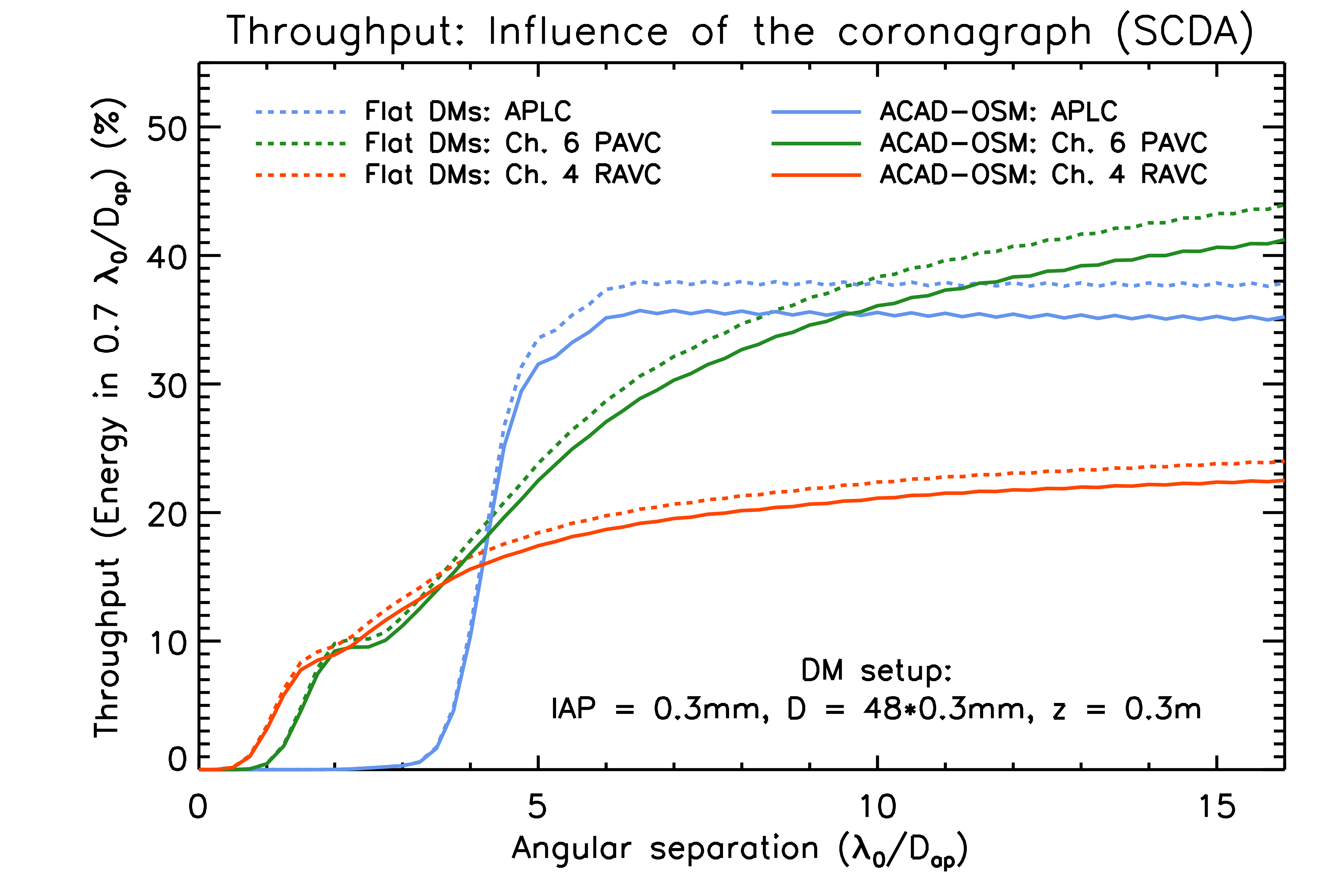}
 \end{center}
\caption[]
{\label{fig:scda_throughput} Throughput performance for the SCDA aperture for an APLC (blue curve), a charge 4 RAVC (red curve) and a charge 6 PAVC (green curve). The dashed lines show the throughput before any correction (due to the coronagraph alone). The solid lines show the throughput at the end of the  correction (with $N_{act} = 48$, IAP = 0.3 mm, D = $48 * 0.3$ mm, $z = 0.3$ m, $\Delta \lambda /\lambda_0 = $ 10\% BW).}
\end{figure}

Fig~\ref{fig:scda_throughput} shows the results in throughput for the same correction parameters. As for all the throughput curves shown in this article, the dashed lines show the throughput before the ACAD-OSM technique, with flat DMs (throughput due to the coronagraph alone). The solid lines show the throughput when the ACAD-OSM shape solutions are applied on the DMs. For each coronagraph, the difference between the two curves (dashed and solid) therefore shows the influence of the ACAD-OSM algorithm on the deformation of the off-axis PSF. In this case, this is limited to a few percents, which shows that the throughput final levels are mainly driven by the coronagraph designs. 

The throughput curve of the APLC shows a steep increase at the edge of the  $4 \lambda_0/D_{ap}$ FPM, followed by a relatively flat throughput. The performance of the two apodized vortex coronagraphs increase with the separation. The charge 4 vortex have a very small advantages at very small separation over the charge 6. However, at large separations, the charge 6 PAVC almost doubles the performance in throughput of the charge 4 RAVC, as expected from \cite{fogarty17}.

In this section, we showed that the contrast floor (for a given DM setup) is a weak function of the static coronagraph solution. The throughput for the planet as a function of separation is however a very strong function of this choice of coronagraph. This is also true for the IWA. The optimization of the coronagraph is therefore critical to obtaining the best performance with ACAD-OSM. We now study the optimization of the coronagraph to obtain the best performance with ACAD-OSM.

%-------------------------------------------------------------------------------------------------
%-------------------------------------------------------------------------------------------------

%-------------------------------------------------------------------------------------------------
%-------------------------------------------------------------------------------------------------

%-------------------------------------------------------------------------------------------------
\section{Optimization of the ACAD-OSM technique for throughput or contrast performance}
\label{sec:fine_tunning}

%-------------------------------------------------------------------------------------------------

\subsection{Context: throughput or contrast optimization ?}

In ACAD-OSM I and the present paper, ACAD-OSM is primarily optimized to maximize the contrast level. However, contrast is not the only metric of interest in the pursuit of the final goal of high contrast imaging: detection of faint companions around nearby stars. \cite{ruane16} showed that for a given set of assumptions for the instrument and observation conditions (e.g. ground- or space-based telescope, size of the primary mirror, amount of aberrations or star and planet fluxes), the estimated integration time to detect a planet is a complex function of both contrast and throughput.

The fine-tuning optimization of the ACAD-OSM technique to maximize planet SNR in each case is out of the scope of this paper. However, in this section, we show how two different parameters can optimize either contrast or throughput at the expense of the other. Using these fine-tuning parameters for a given coronagraphic instrument, one can achieve the best performance for a specific instrument. {These parameters are especially useful in the case where ACAD-OSM has an important effect on the throughput, \textit{i.e.} in the case of large struts, as shown in ACAD-OSM I, Sec. 4.2. For this reason, in this section, the WFIRST-like aperture is studied, with a 36\% central obscuration and large struts (Figure~\ref{fig:wfirst_ls46_optimdm_dh}, top, left).}

%-------------------------------------------------------------------------------------------------
\subsection{Size of the Lyot stop radius with the WFIRST aperture}
\label{sec:LSradius}
%-------------------------------------------------------------------------------------------------

Some coronagraph designs (APLC, PAVC) provide the possibility of choosing the inner and outer radius of the Lyot stop and to optimize the apodization for this Lyot Stop. In \cite{ndiaye15} and \cite{fogarty17}, the authors explore the parameter space of these parameters to maximize the throughput in the final focal plane. Since the optimization of this parameter for each coronagraph has been done in previous papers \cite{ndiaye16} and \cite{fogarty17}, we only focus of its impact on the ACAD-OSM method performance.

\begin{figure}
\begin{center}
 \includegraphics[trim= 1.5cm 0.8cm 1.0cm 0.5cm, clip = true,width = .48\textwidth]{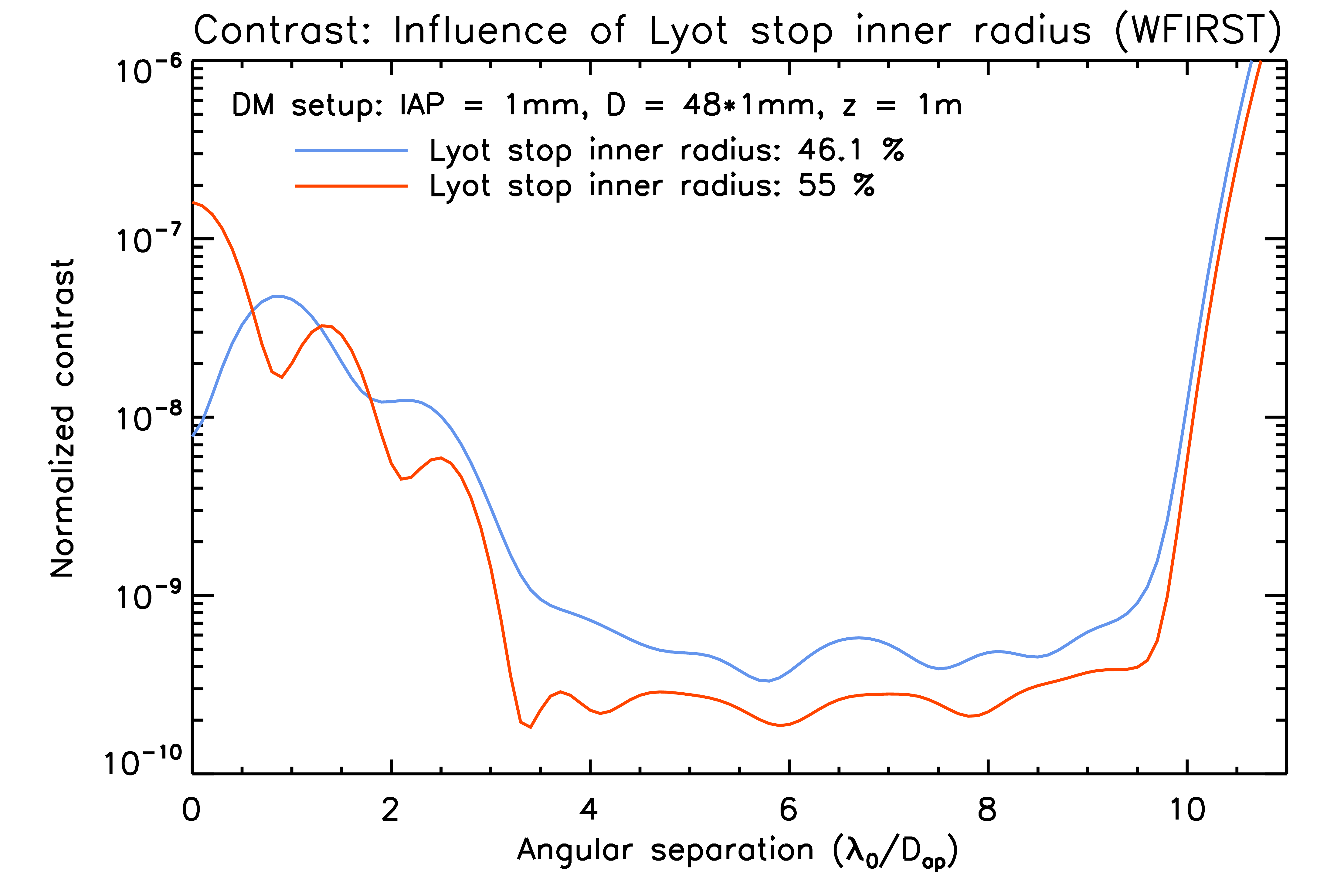}
  \end{center}
\caption[]
{\label{fig:sizeLS_contrast} Influence of the Lyot stop inner radius on the contrast performance. The WFIRST aperture and DM setup are used ($N_{act} = 48$, IAP = 1 mm, D = $48 * 1$ mm, $z = 1$ m, $\Delta \lambda /\lambda_0 = $ 10\% BW)  to show this effect. The correction with an inner radius of the Lyot stop of 46.1\% is shown in blue and with a larger inner radius of the Lyot stop (55\%) in red.}
\end{figure}

\begin{figure}
\begin{center}
 \includegraphics[trim= 2.5cm 0.8cm 1.0cm 0.5cm, clip = true,width = .48\textwidth]{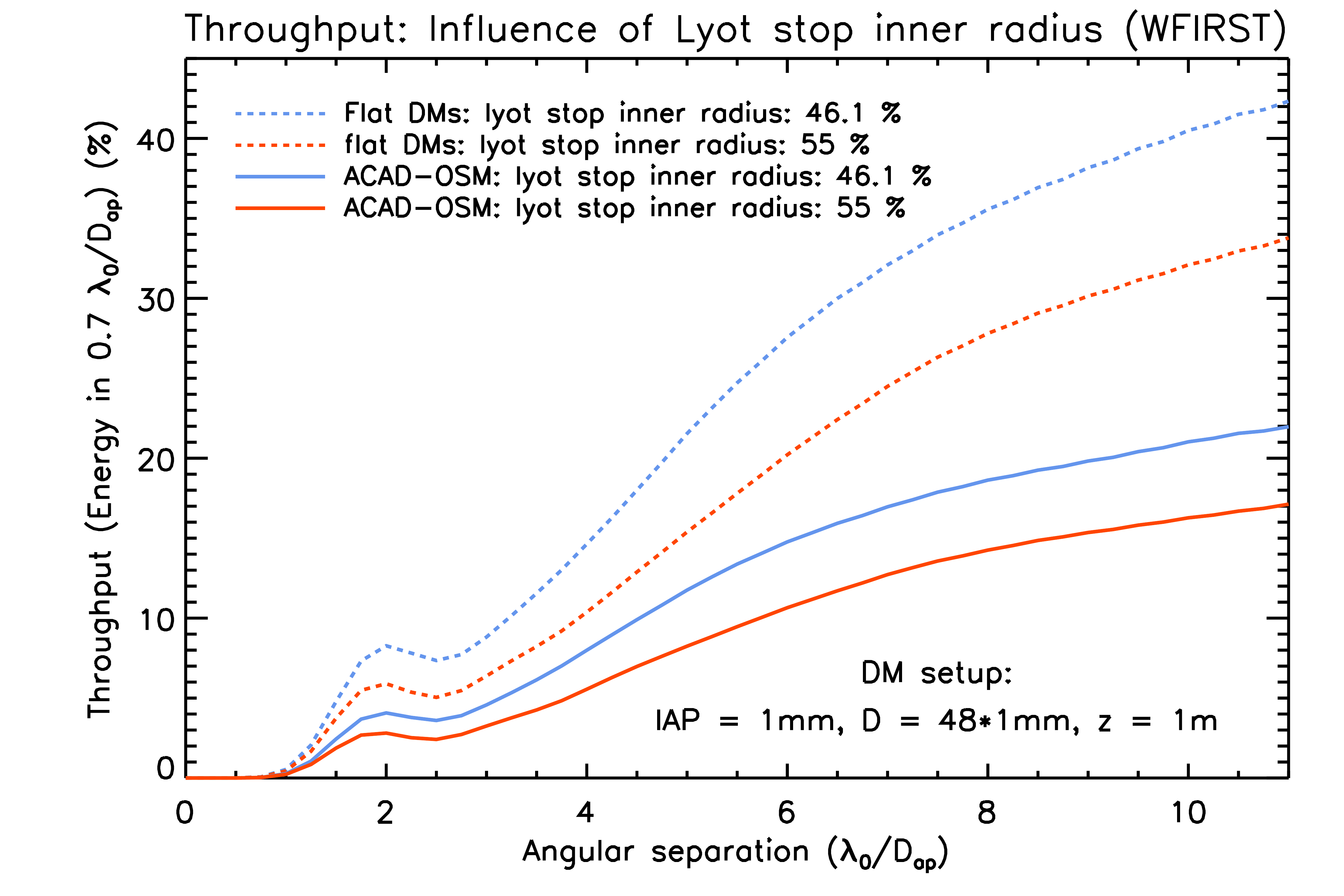}
 \end{center}
\caption[]
{\label{fig:sizeLS_throughput} Influence of the Lyot stop inner radius on the throughput performance, with the WFIRST aperture and DM setup ($N_{act} = 48$, IAP = 1 mm, D = $48 * 1$ mm, $z = 1$ m, $\Delta \lambda /\lambda_0 = $ 10\% BW). The dashed lines show the throughput before any correction (due to the PAVC charge 6 alone). The correction with an inner radius of the Lyot stop of 46.1\% is shown in blue and with a larger inner radius of the Lyot stop (55\%) in red.}
\end{figure}

In this section, the WFIRST DM setup is used ($N_{act} = 48$, IAP = 1 mm, D = $48 * 1$ mm, $z = 1$ m, $\Delta \lambda /\lambda_0 = $ 10\% BW. Two different charge 6 PAVCs are compared, which have both been optimized for this 36\% central obscuration but using two different Lyot inner radii: 46.1\% (which is normally the optimum for energy transmission for a 36\% central obscuration aperture) and 55\%. The results in contrast are presented in Figure \ref{fig:sizeLS_contrast} and in throughput performance in Figure \ref{fig:sizeLS_throughput}. The dashed lines show the throughput before any correction (due to the PAVC charge 6 alone). The solid lines show the throughput at the end of the  ACAD-OSM correction. For both contrast and throughput performance, the smaller inner radius of the Lyot stop (46.1\%) is represented with blue lines and the 55\% inner radius Lyot stop is represented with red lines.

A larger Lyot stop inner radius has better contrast performance but worse throughput performance. The better performance in contrast is due to the fact that the parts of the aperture hidden behind the Lyot stop have a lot less impact on the focal plane and are easier to correct for by the DMs. For this reason, a larger Lyot stop inner radius relieves the DMs and helps them achieve deeper contrast level. However, a larger Lyot stop inner radius often degrades the throughput of the coronagraph itself (Fig. \ref{fig:sizeLS_throughput}, dashed lines), since more off-axis light is blocked.

%-------------------------------------------------------------------------------------------------
%-------------------------------------------------------------------------------------------------

%-------------------------------------------------------------------------------------------------
%-------------------------------------------------------------------------------------------------

The Lyot stop inner radius can therefore be used as a fine-tuning parameter which can allow us to increase either performance in throughput or contrast at the expense of the other. For most of this paper, the coronagraph with a 55\% inner radius Lyot stop will be used for the WFIRST aperture, allowing us to almost reach the symbolic $10^{-10}$ contrast level with this DM setup. However, this good performance in contrast come at the expense of throughput.

\subsection{Number of matrices in the ACAD-OSM algorithm}
\label{sec:Matrix_number}

Figure~\ref{fig:sizeLS_throughput} shows that the planet throughput is both impacted by the choice of Lyot Stop (difference between the two dashed curves) and by the PSF distortion due to the ACAD-OSM shapes (the difference between the two blue curves is more important that the difference between the two red curves). This warrants the question of whether or not there exist ACAD-OSM shapes that still meet contrast requirements but yield better throughput \citep[and thus improve planet SNR, as explained in][]{ruane15}.

In ACAD-OSM I paper, we explained that this algorithm is based on the search for a nearby contrast minimum. To achieve this goal, the interaction matrix is recomputed several times to avoid being limited by the linearity range allowed by the stroke minimization algorithm around the initial DM shapes. We showed that after a certain number of matrices, the improvement of contrast achieved with a new matrix is minimal, because convergence is no longer limited by linearity but by proximity to a local minimum in contrast. We noted that for a $\Delta \lambda /\lambda_0 = $ 10\% BW, 8 matrices were usually enough to achieve the local minimum allowed by the coronagraph and DM setup. However, one can decide to not push the algorithm to its limit and stop before reaching the best contrast point. This point is called the operating point of ACAD-OSM.

\begin{figure}
\begin{center}
 \includegraphics[trim= 2.5cm 0.8cm 1.0cm 0.5cm, clip = true,width = .48\textwidth]{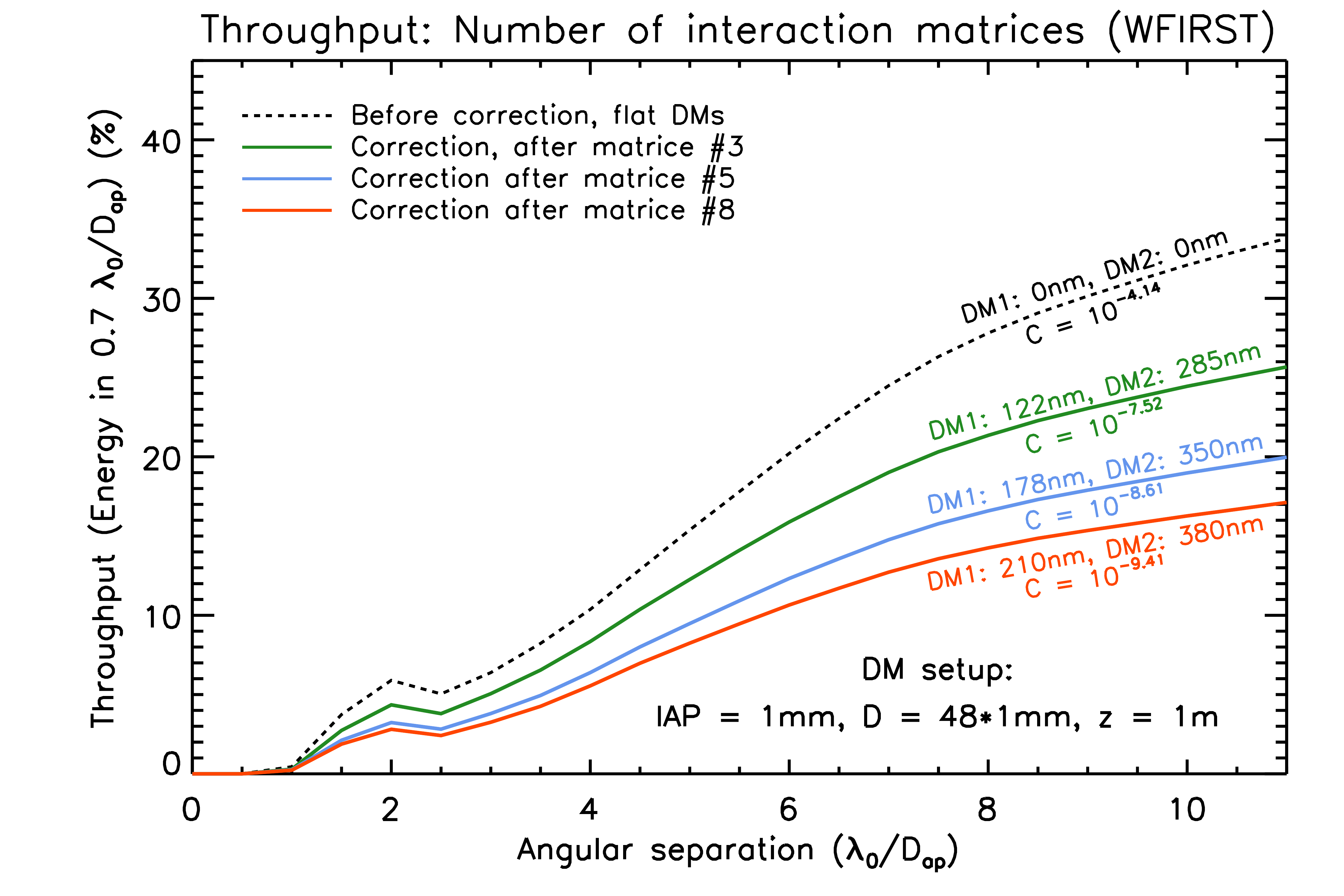}
 \end{center}
\caption[]
{\label{fig:numberMat_throughput} Influence of the number of interaction matrices on the throughput performance with a charge 6 PAVC the WFIRST aperture and DM setup ($N_{act} = 48$, IAP = 1 mm, D = $48 * 1$ mm, $z = 1$ m, $\Delta \lambda /\lambda_0 = $ 10\% BW). The dashed line shows the throughput before any correction (due to the PAVC charge 6 alone). The solid lines show the throughput after the correction with matrix \#3, \#5 and \#8. }
\end{figure}

Fig.~\ref{fig:numberMat_throughput} shows the throughput level after several steps of the ACAD-OSM correction. The red solid curve corresponds to the results after the 8th matrix, probably close to the local contrast minimum. In each case the strokes on the DMs (peak-to-valley, PV) and the contrast level reached in the 3-10 $\lambda_0/D_{ap}$ DH are printed . The closer the correction get to the local contrast minimum, the more the strokes increase and the more the throughput degrades. This shows that the step at which the correction is stopped provides us with another fine-tuning parameter which, for a given DM setup and a given coronagraph, can allow us to trade between throughput and contrast.

%-------------------------------------------------------------------------------------------------
%-------------------------------------------------------------------------------------------------

\section{Robustness to small TT jitter}
\label{sec:jitter}

\begin{figure}
\begin{center}
 \includegraphics[trim= 1.5cm 0.8cm 1.0cm 0.5cm, clip = true,width = .48\textwidth]{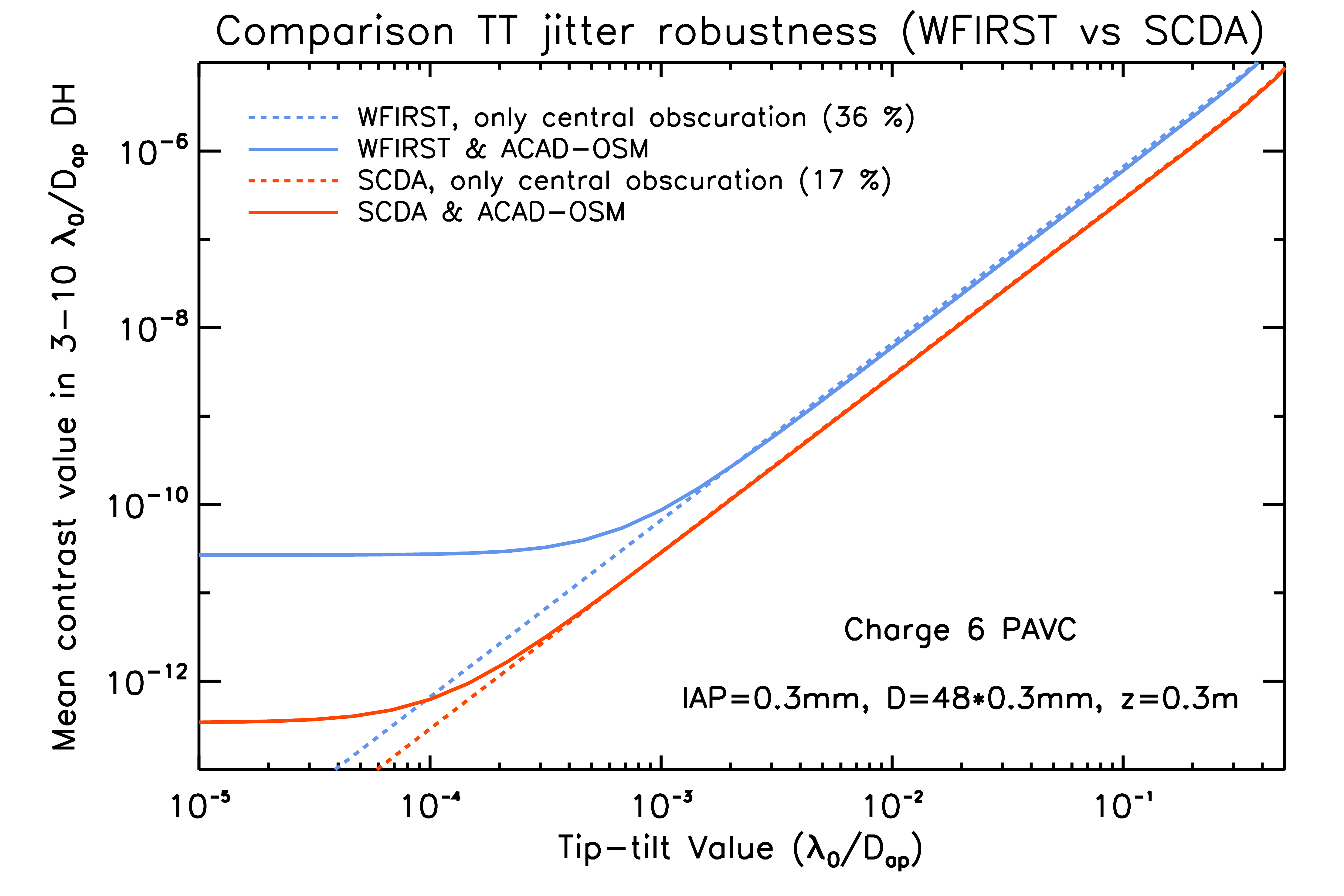}
 \end{center}
\caption[]
{\label{fig:wfirst_scda_throughput_compar} TT jitter robustness for two apertures, with the same coronagraph (Charge 6 PAVC), the same DM setup  ($N_{act} = 48$, IAP = 0.3 mm, D = $48 * 0.3$ mm, $z = 0.3$ m, $\Delta \lambda /\lambda_0 = $ 10\% BW) and the same DH (3-10 $\lambda_0/D_{ap}$)}
\end{figure}

{The robustness to small low order aberrations is a major concern of the current generation of coronagraph design. In particular, the impact on the performance of small TT jitter, which is equivalent problem to the impact of resolved stars, will be a major limitation of future large telescopes. Indeed, future large space-based telescopes will typically observe stars that are partially resolved, where the stellar angular diameter is approximately a tenth of the diffraction limit. In this section, we study the robustness to small TT jitter of the ACAD-OSM solutions. The same coronagraph (charge 6 PAVC) is used, for 2 different apertures (WFIRST aperture and SCDA aperture). A favorable DM setup ($N_{act} = 48$, IAP = 0.3 mm, D = $48 * 0.3$ mm, $z = 0.3$ m, $\Delta \lambda /\lambda_0 = $ 10\% BW) is chosen: the impact of this parameter on this metric will be studied in detail in Section~\ref{sec:jitter_fresnel}. Fig~\ref{fig:wfirst_scda_throughput_compar} shows the impact of TT jitter on the average contrast level in the 3-10 $\lambda_0/D_{ap}$ DH for these two aperture. The dashed line shows the performance of the coronagraph itself (just a central obscuration and the coronagraph), and the solid line shows the performance of the coronagraph and ACAD-OSM with the full aperture. }

{The robustness of the coronagraphs themselves to TT jitter is not studied in this paper. None of the coronagraphs of this paper have been optimized for robustness to TT, only for throughput. However, this parameter has been studied in the case of the PAVC \citep{fogarty17SPIE}, and in the case of the APLC \citep{ndiaye16}. In the case of the current version of the PAVC, the two coronagraph robustness is mainly a function of the size of the central obscuration, as shown by the difference between the two dashed lines.}

%-------------------------------------------------------------------------------------------------
%-------------------------------------------------------------------------------------------------

{For small TT jitters, the contrast level is better for the coronagraph alone than for the full aperture and ACAD-OSM correction. A better contrast can be achieved with the SCDA aperture, due to the finer struts. When the TT jitter robustness of the coronagraph itself reaches the level of the DH, the full aperture and ACAD-OSM correction contrast starts to degrades. For larger TT jitters, the contrast is entirely driven by the TT robustness of the coronagraph itself.}

%-------------------------------------------------------------------------------------------------
%-------------------------------------------------------------------------------------------------
%-------------------------------------------------------------------------------------------------

\section{Impact of DM setup: analytic formalism}
\label{sec:formalism}
%-------------------------------------------------------------------------------------------------
%-------------------------------------------------------------------------------------------------
%-------------------------------------------------------------------------------------------------

\subsection{Context}

As discussed in the introduction, the other parameter that influences performance is the DM setup (size of the beam, IAP and distance between DMs). In this section, we describe a simple formalism for the propagation between mirrors. This formalism has been developed to understand the limitations of the two DM aperture discontinuity correction, but is not specific to the ACAD-OSM correction algorithm and can be applied to any two DM correction algorithms, such as the ones used for correction of phase and amplitude aberrations \citep{pueyo09, beaulieu17} or the active technique developed for the WFIRST instrument \citep{krist16}. It can also be useful for fixed mirror apodization techniques \citep{guyon05,Guyon14,fogarty17}, as long as they introduce small enough apodization \citep[as defined in ][]{mazoyer16b} and can therefore be simulated using the Fresnel assumptions.

%-------------------------------------------------------------------------------------------------
\subsection{Size of the beam in the second DM plane}
\label{sec:oversize_desc}
%-------------------------------------------------------------------------------------------------

\begin{figure}
\begin{center}
 \includegraphics[trim= 0.5cm 0.5cm 0.7cm 0.2cm, clip = true,width = .48\textwidth]{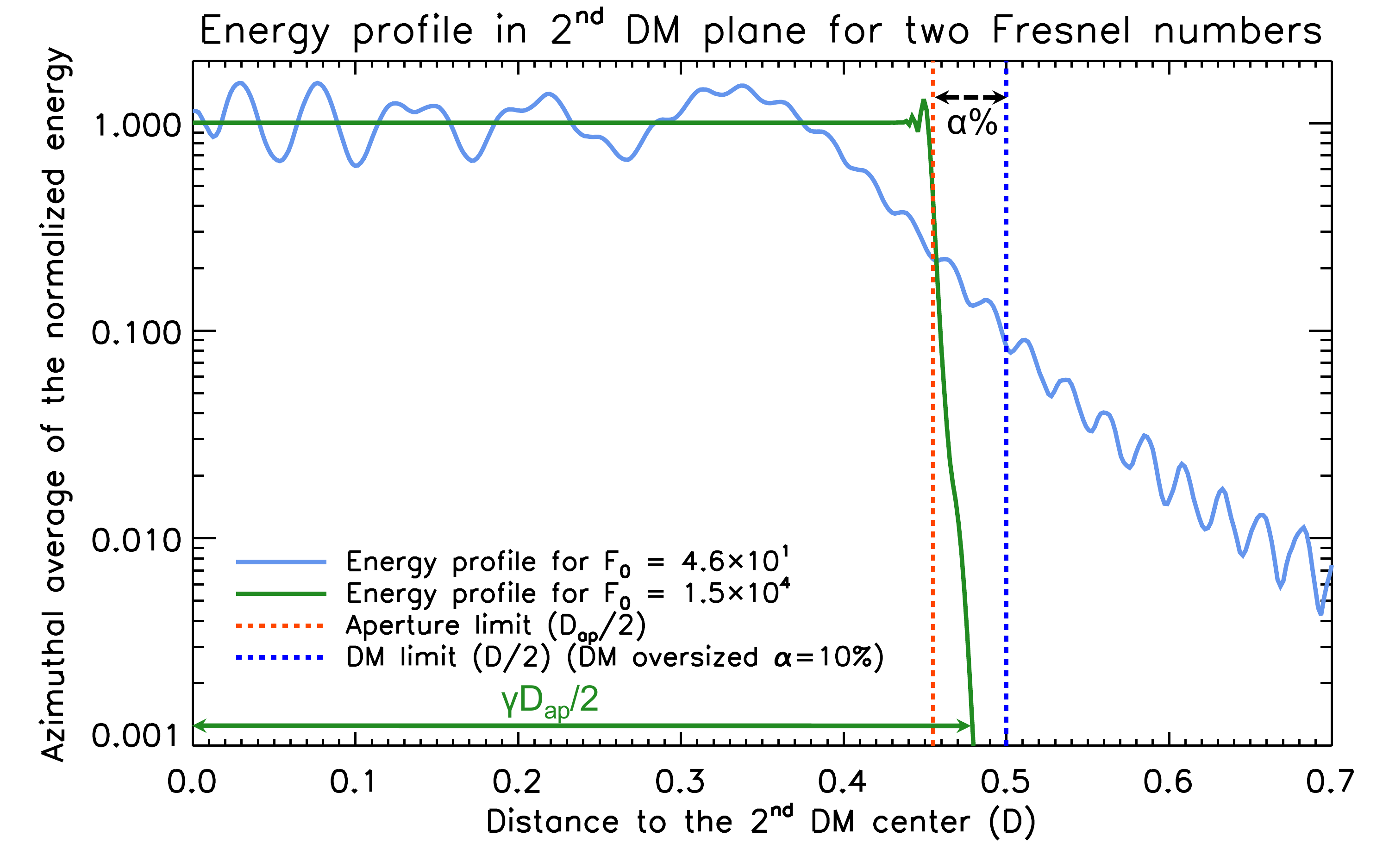}
 \end{center}
\caption[plop]
{\label{fig:oversizing_azim} Azimuthal average of the normalized diffracted energy in second DM plane for  $F_0= 4.6\times 10^{1}$ (blue solid line) and $F_0= 1.5 \times 10^{4}$ (green solid line). The vertical dashed lines show the limit of the aperture (in red) and of the DM (in blue).}
\end{figure}

We recall that the first DM is a square of size $D \times D$ and is in the pupil plane of the circular aperture, of diameter $D_{ap}$ (see Fig.~\ref{fig:schema_ACAD}). The DMs are slightly oversized compared to the pupil:
\begin{equation}
\label{eq:alphadef}
D = (1+\alpha)D_{ap}\,\,\,\,\,\,\,\,\, \alpha \ge 0
\end{equation}
In practice, $\alpha$ is expressed as a percentage (from 0\% to 30\%). The second DM, also a square of size $D \times D$, is located at a distance $z$ from the first DM. Due to the Fresnel propagation, the diameter of the beam expands between the first and second DM. The diameter of the beam in the second DM plane is labeled $D_{ap, 2^{nd}DM}$. We define $\gamma$ the relative size of this aperture compared to the initial aperture:
\begin{equation}
D_{ap, 2^{nd}DM} = \gamma D_{ap}\,\,\,\,\,\,\,\,\, \gamma > 1
\end{equation}
Finally, the propagation parameter of this setup is called the Fresnel number:
\begin{equation}
\label{eq:fnum}
F_0 = \dfrac{D^2}{\lambda_0 z}.
\end{equation}
Fig.~\ref{fig:oversizing_azim} shows the effects of different Fresnel numbers on the diffraction in the second DM plane. The solid blue and green lines show the azimuthal average of the normalized diffracted energy in this plane for two Fresnel numbers. The vertical dashed lines show the limit of the aperture ($D_{ap}/2$, in red) and of the DM ($D/2 = (1 +\alpha) D_{ap}/2$, in blue). The parameter $\alpha$ and $\gamma$ are shown in the figure. 

In both cases, the images of the diffracted apertures are larger than the original aperture size. However, in the large Fresnel number case (green solid line), the energy mostly fits in the oversized DM, while in the small Fresnel number case (blue solid line) a non-negligible portion of the energy leaks outside of the DM. Both cases show ringing in the second DM plane. {These rings, called Gibbs rings, are very chromatic and have been described in details in} \cite{pueyo07}.

%-------------------------------------------------------------------------------------------------
%-------------------------------------------------------------------------------------------------

\begin{figure}
\begin{center}
 \includegraphics[trim= 1.5cm 0.9cm 1.0cm 0.5cm, clip = true, width = .48\textwidth]{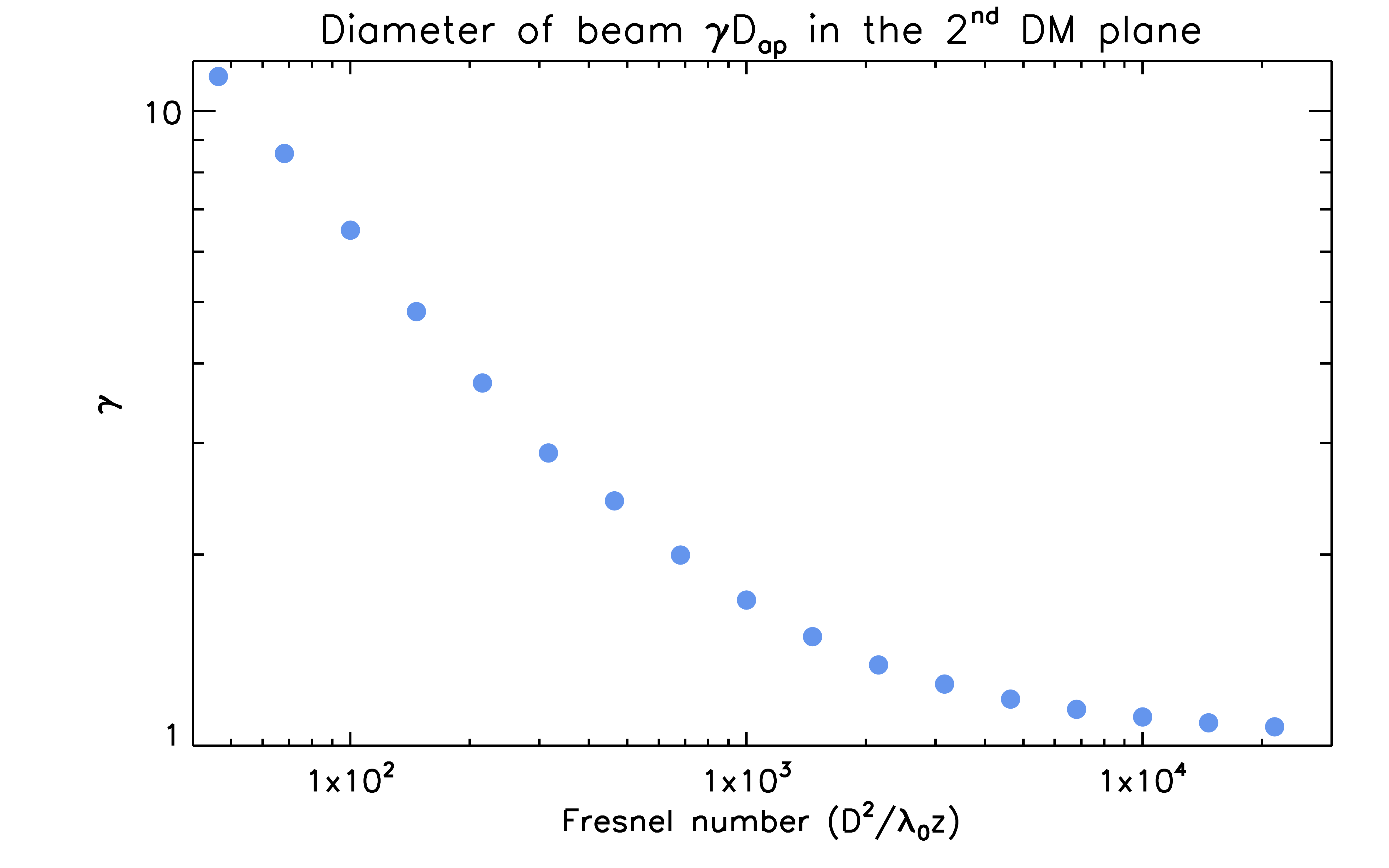}
 \end{center}
\caption[plop]
{\label{fig:oversizing_size} Ratio $\gamma$ (of the diameter of the diffracted aperture in the second DM plane compared to the original aperture diameter) as function of the Fresnel number $F_0$.}
\end{figure}

We now analyze the relationship between the extent of the diffracted beam in the second DM plane and the Fresnel number. We numerically simulate the Fresnel propagation of a clear aperture in seventeen Fresnel number cases equally distributed in log scale from $F_0 = 4 \times 10^{1}$ to $F_0 = 2 \times 10^{4}$. We measure the diameter of the diffracted aperture in the second DM plane (measured by setting a $10^{-5}$ threshold on the azimuthal average of the energy) and compare it to the original aperture radius to obtain the stretching $\gamma$. The results are plotted in blue dots in Fig.~\ref{fig:oversizing_size}. 

%-------------------------------------------------------------------------------------------------
%-------------------------------------------------------------------------------------------------

It shows that the diameter of the beam on the second DM can be as large as 10 times the diameter of the original aperture for small Fresnel numbers. This is much larger than the second DM size, which introduces significant vignetting of the beam, studied in the next section.

%-------------------------------------------------------------------------------------------------
\subsection{Impact of the vignetting (small-Fresnel number effect)}
\label{sec:oversize_vigneting}
%-------------------------------------------------------------------------------------------------

%-------------------------------------------------------------------------------------------------
\begin{figure}
\begin{center}
 \includegraphics[trim= 0.8cm 1cm 1.8cm 0.2cm, clip = true,width = .48\textwidth]{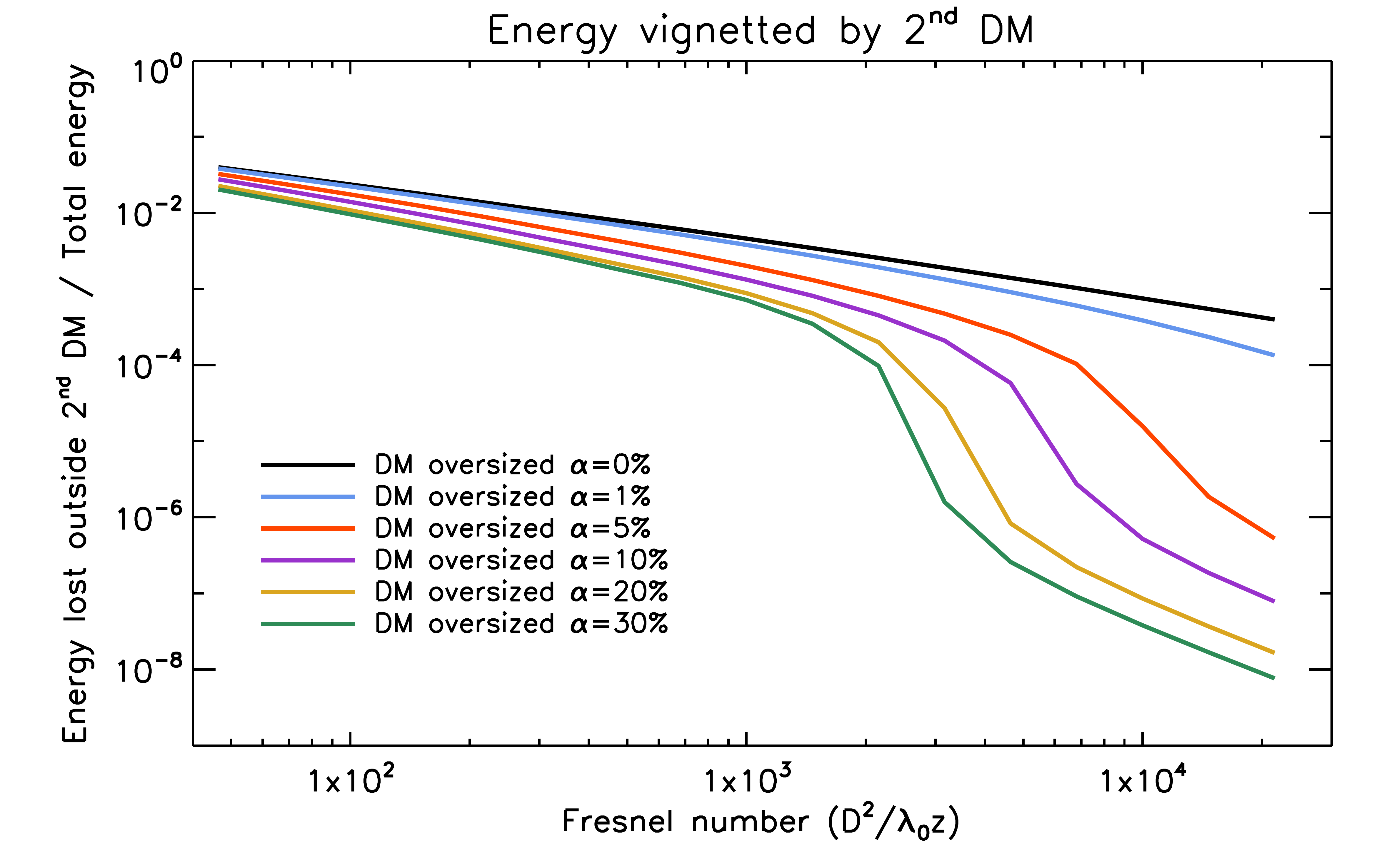}
 \end{center}
\caption[plop]
{\label{fig:oversizing_energy} Ratio of the amount of energy lost due to the vignetting of the second DM on the total energy in the aperture, as function of the Fresnel number for several oversizing cases.}
\end{figure}
%-------------------------------------------------------------------------------------------------

%-------------------------------------------------------------------------------------------------
\subsubsection{Impact on the on-axis transmission}
\label{sec:vignetting_transmission}
%-------------------------------------------------------------------------------------------------

If the energy going outside of the second DM is vignetted and lost, this has an impact on the transmission of the 2 DM system. Indeed, we showed in the previous section that the beam can expands up to 10 times the diameter of the aperture for small Fresnel numbers. However, the effect on on-axis transmission is often negligible. 

Fig.~\ref{fig:oversizing_energy} shows the ratio of the amount of energy lost due to the vignetting of the second DM to the total energy in the aperture as function of the Fresnel number for a range of the oversizing parameter $\alpha$ (from 0 to 30\%). The curves shown are approximated by two straight lines separated by a step, corresponding to the moment when the second DM starts to crop the aperture. However, even in the worst cases ($F_0< 10^{2}$ and $\alpha = 0\%$), the loss of energy is limited to a few percents of the total energy. This means that even if the aperture is more and more diffracted as the Fresnel number decreases, most of the energy still remains inside of the original size of the aperture. 

The on-axis transmission in not the most critical value affecting the performance of a coronagraph. In the next section, we study the impact of vignetting on off-axis throughput. 

%-------------------------------------------------------------------------------------------------
\subsubsection{Impact on the off-axis transmission and throughput}
\label{sec:vignetting_throughput}
%-------------------------------------------------------------------------------------------------

In this section, we analyze the effect of the Fresnel number on the throughput of an off-axis planet located at an angular separation of $n\lambda_0/D_{ap}$ from the on-axis star. A geometrical argument is first used to derive an theoretical law linking the off-axis transmission of the system with the Fresnel number. Then, numerical simulations show the precise impact of this parameter on the performance in off-axis throughput ({energy in the PSF core}).

\paragraph{Geometric analysis of the off axis-transmission}

\begin{figure}
\begin{center}
 \includegraphics[trim= 0.7cm 0cm 0cm 0.5cm, clip = true,width = .48\textwidth]{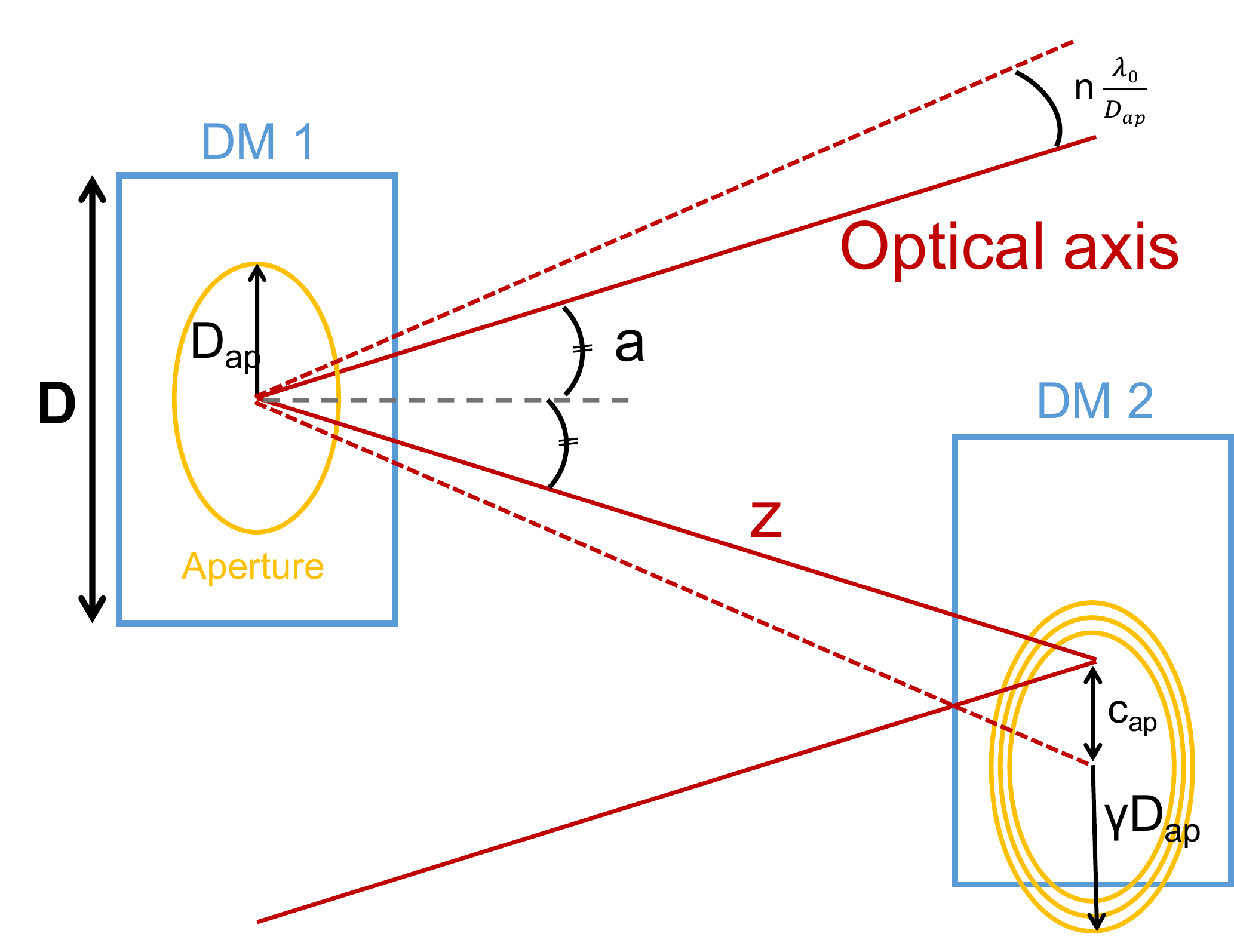}
 \end{center}
\caption[plop]
{\label{fig:schema_throughputlossgeometry} Geometric schematic showing the vignetting of the energy of an off-axis planet located at $n\lambda_0/D_{ap}$ of the star.}
\end{figure}

Fig.~\ref{fig:schema_throughputlossgeometry}, shows that the distance $c_{ap}$ from the center of the aperture image to the center of the second DM is the angle times the distance between the DMs: 
\begin{equation}
    c_{ap} \simeq   \dfrac{n \lambda_0}{D_{ap}} z  =  n\dfrac{(1+\alpha) \lambda_0 z}{D} \,\,,
\end{equation}
using Eq~\ref{eq:alphadef}. a is the angle between the DMs. In all this paper, $a = 0$ for simplicity. Assuming flat DMs and an axisymmetric aperture for simplicity, the energy in the beam has a radial symmetry. In that case, half of the diffracted beam is outside of the second DM when $c_{ap}(n) = D/2$, which corresponds to a loss of 50\% of the transmission. We call this separation $n_{\textrm{Tr}50\%}$:

\begin{equation}
\label{eq:n50}
n_{\textrm{Tr}50\%} =  \dfrac{1}{2 (1+\alpha)}\dfrac{D^2}{\lambda_0 z} = \dfrac{1}{2 (1+\alpha)}F_0\,\,.
\end{equation}
This simple analysis shows that this effect is linearly dependent on the Fresnel number and favors large values of $F_0$. However, this analytic formula is not a practical tool for designing future two DM coronagraphic instruments: a 50\% loss of the off-axis energy transmission, without even considering the impact of the vignetting on the shape of the PSF, is not acceptable.

\paragraph{Numerical analysis of the off axis-throughput ({energy in the PSF core})}

\begin{figure}
\begin{center}
 \includegraphics[trim= 0.5cm 0.5cm 0.7cm 0.2cm, clip = true,width = .48\textwidth]{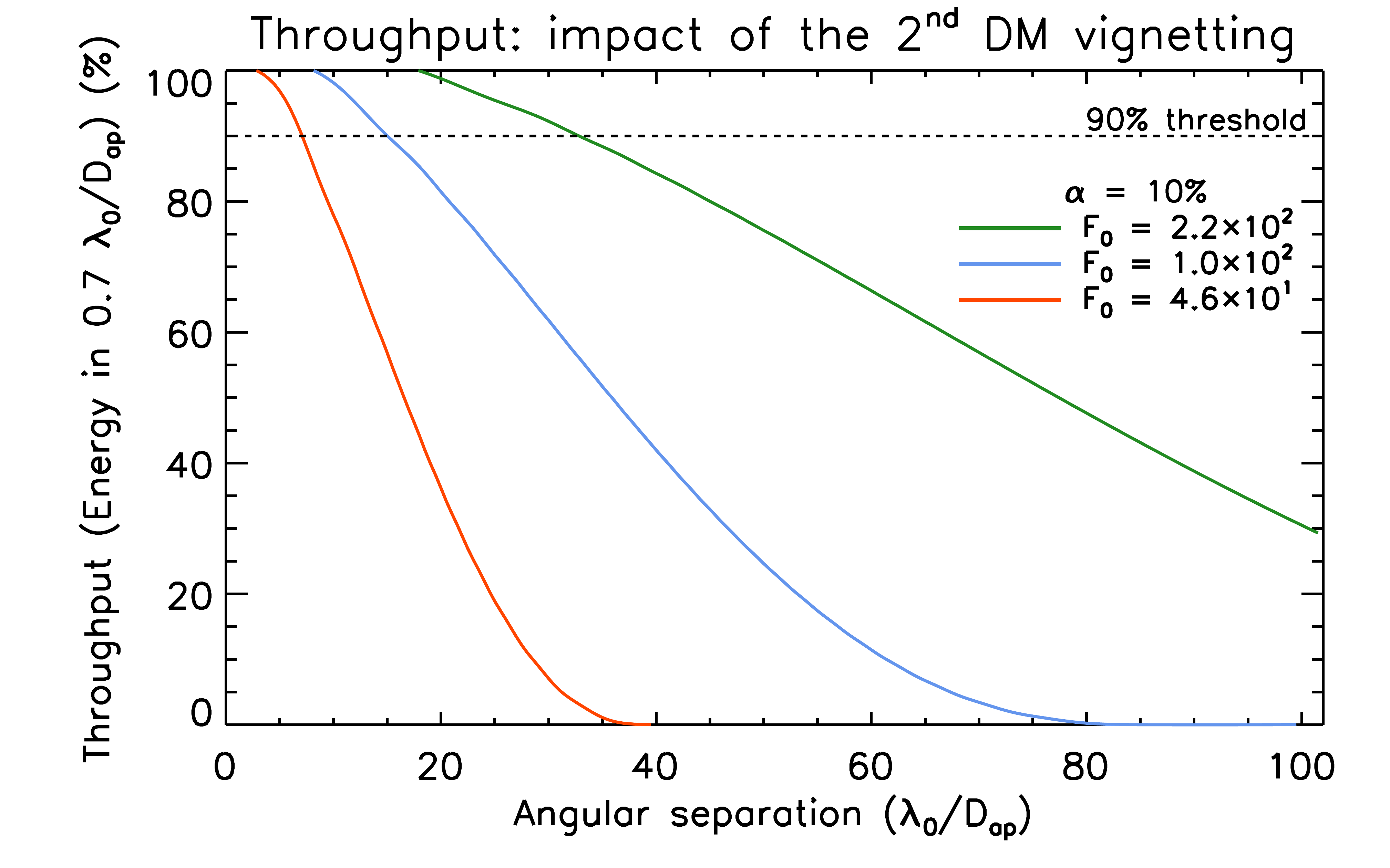}
 \end{center}
\caption[plop]
{\label{fig:throughput_funtt_fnum} Throughput ({energy in the PSF core}) loss due to the second DM vignetting for 3 Fresnel numbers and $\alpha$ = 10\%.}
\end{figure}

{The transmission of the 2 DM system (total light captured by DM2) is not a useful metric and it is more useful to discuss how much light is actually contained in the PSF core.} A simple numerical simulation can be used to estimate the loss of off-axis throughput ({energy in the PSF core}) due to the vignetting of the second DM. From an initial clear aperture, we measure the loss of energy in the PSF core (throughput) in the next focal plane (with no coronagraph). Fig~\ref{fig:throughput_funtt_fnum} shows the throughput at as a function of separation for three Fresnel numbers and $\alpha$ = 10\%. The 90\% throughput threshold (10\% throughput loss) is represented with a dashed horizontal line. 

\begin{figure}
\begin{center}
  \includegraphics[trim= 1.8cm 1cm 1.8cm 0.5cm,width = .48\textwidth]{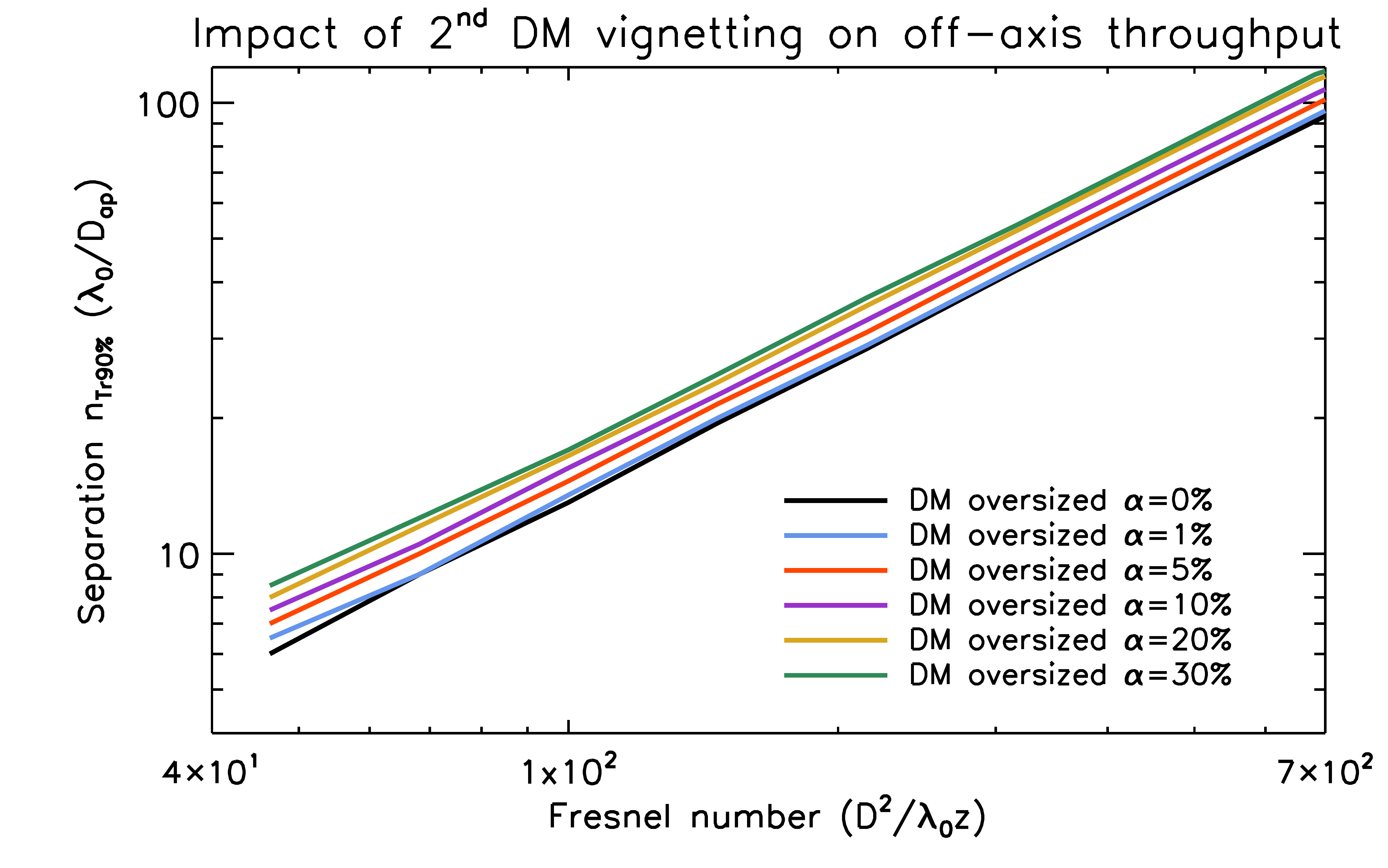}
 \end{center}
\caption[plop]
{\label{fig:oversizing_throughput} These curves represent $n_{\textrm{Th}90\%}$, the separation at which the two DM system throughput ({energy in the PSF core}) is only 90\%, due to the vignetting of the second DM, as function of the Fresnel number for several oversizing cases.}
\end{figure}

We call $n_{\textrm{Th}90\%}$ the smallest angular separation for which the throughput of the two DM system drops under 90\%. To understand the impact of the vignetting of the second DM for all DM setups, this simple numerical simulation is repeated for several Fresnel numbers and several over-sizing cases ($\alpha$ from 0 to 30 \%). The results are plotted in Fig.~\ref{fig:oversizing_throughput}. Only the results for $F_0< 7\times10^{2}$ are plotted. 

%%%-----------------------------------------------------------------------------------------------------
%%%-----------------------------------------------------------------------------------------------------

%-------------------------------------------------------------------------------------------------
%-------------------------------------------------------------------------------------------------

%%%-----------------------------------------------------------------------------------------------------
%%%-----------------------------------------------------------------------------------------------------
These curves also shows the linear trend predicted by Eq.~\ref{eq:n50} {for the transmission of the system}. This effect in practice prevents the use of any system with $F_0< 2\times10^{2}$, corresponding to $n_{\textrm{Th}90\%} < 30 \lambda_0/D_{ap}$, or even further depending on the aimed-for DH size. This effect might have an impact on the maximum OWA achievable for a given Fresnel number. However, if $n_{\textrm{Th}90\%} \gg  N_{act}/2$ this effect is negligible and the limiting factor for the OWA remains the number of actuators on the DM. This two DM system throughput analysis, with flat mirrors and without any coronagraph, does not take into account the throughput of the coronagraph, nor the off-axis throughput loss due to the ACAD-OSM shapes on the DMs, which is studied independently in Section~\ref{sec:DM_setup}.

In the rest of this article, we assume a setup where the second DM is surrounded by a non-actuated reflective surface that extends the side length of the second mirror to twice the length of the DM. This allows us to ignore the two DM system vignetting effects on throughput described in this section. Most of the throughput performance shown in this paper are therefore only due to either the coronagraph or the ACAD-OSM shapes on the DMs.

%-------------------------------------------------------------------------------------------------
\subsubsection{Impact on the contrast}
\label{sec:vignetting_contrast}
%-------------------------------------------------------------------------------------------------

{A sharply vignetted beam in the second DM plane can create diffraction leaks in the focal plane of the coronagraph that eventually limits the contrast. The ACAD-OSM algorithm is aiming at minimizing the contrast in the DH and would naturally converges towards a solution limiting this effect, if it becomes the limiting factor on the contrast level. However, the Gibbs rings shown in Fig~\ref{fig:oversizing_azim} are very chromatic \citep{pueyo07} and therefore, for small Fresnel numbers, it becomes impossible to correct for this effect on a large bandwidth.}

%%%-----------------------------------------------------------------------------------------------------
\begin{figure*}
\begin{center}
   \includegraphics[trim= 1.8cm 1cm 1.8cm 0.5cm,width = .48\textwidth]{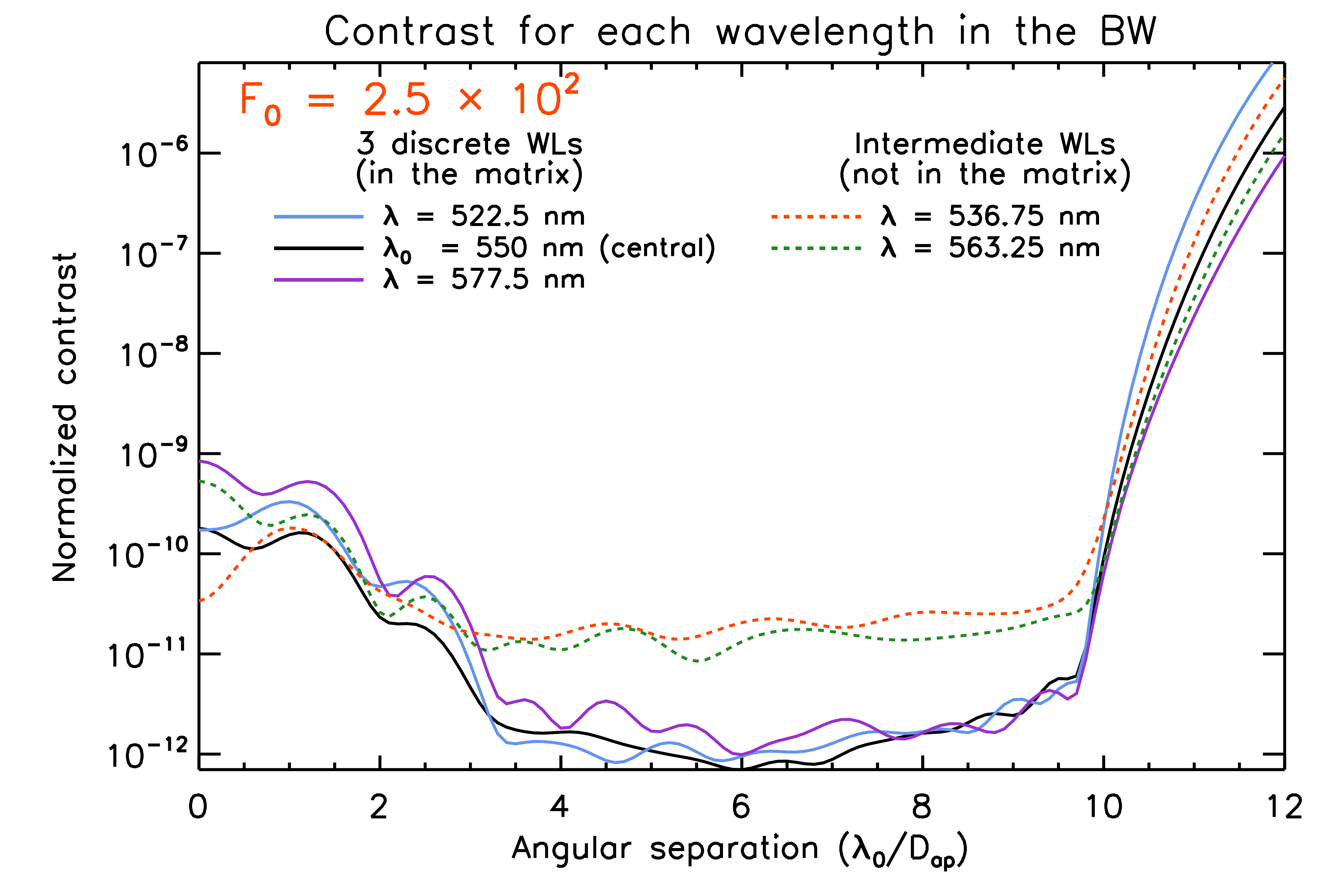}
  \includegraphics[trim= 1.8cm 1cm 1.8cm 0.5cm,width = .48\textwidth]{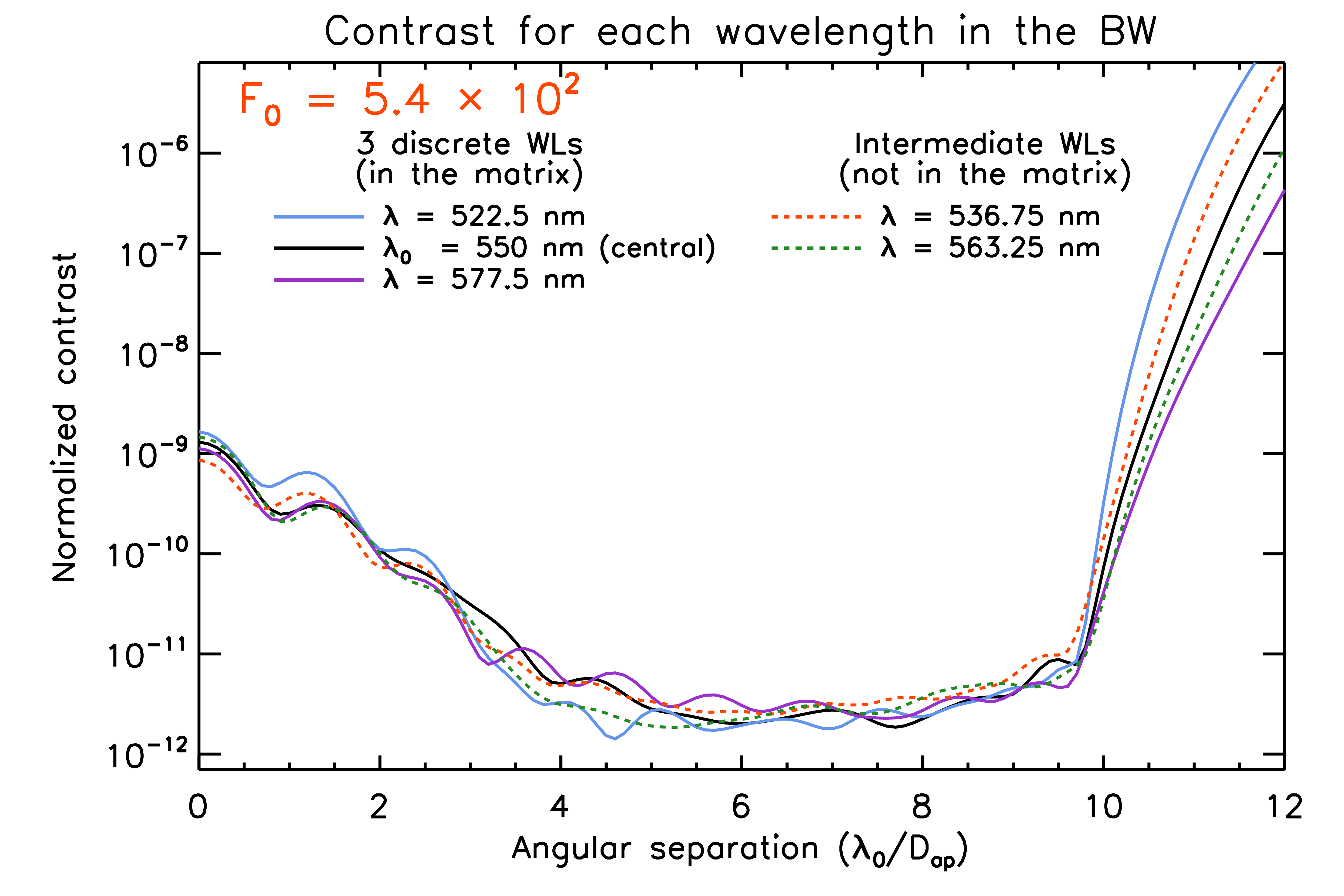}
 \end{center}
\caption[plop]
{{\label{fig:contrast_vignetting} Impact on the vignetting on the contrast. Contrast curves for discrete wavelengths in the BW for two Fresnel numbers. In this numerical simulation, $\alpha = 10\%$ and the second DM is surrounded by a non-actuated reflective surface that extends the side length of the second mirror to twice the length of the DM.}}
\end{figure*}
%%%-----------------------------------------------------------------------------------------------------
{Fig.~\ref{fig:contrast_vignetting} shows using numerical simulations, the limitation of the ACAD-OSM in this case, with the WFIRST aperture. In this numerical simulation, $\alpha = 10\%$ and the second DM is surrounded by a non-actuated reflective surface that extends the side length of the second mirror to twice the length of the DM, which means that this effect only impacts the correction when the diffracted aperture is larger than $2.2\, D_{ap}$ ($\gamma > 2.2$). Fig.~\ref{fig:oversizing_size} shows that it corresponds to $F_0< 5\times 10^2$. To understand the impact on the correction, ACAD-OSM correction were made at $F_0= 2.5\times 10^2$ and $F_0= 5.4\times 10^2$ with a 10\% BW. }

{We recall that in the ACAD-OSM algorithm, the contrast is first obtained with 3 discrete wavelengths in the interaction matrix, and then applied to a more continuous BW to measure results. Once the DM shapes were obtained using these three wavelengths, the contrast achieved after the correction for several discrete wavelengths in that BW is plotted in Fig.~\ref{fig:contrast_vignetting}. The solid lines are the contrast curves for the three discrete wavelengths used in the interaction matrix (the central and extreme wavelengths), the two dashed lines are contrast curves for two intermediary wavelengths. For $F_0= 5.4\times 10^2$ (right plot) the Gibbs rings have not reach the edge of the second DM and the correction obtained with only three discrete wavelengths (solid lines) also works for the intermediate wavelengths (dashed lines). This is clearly not the case for a smaller Fresnel number ($F_0= 2.5\times 10^2$, left plot), when the Gibbs rings have reached the edge of the second DM.}

{This effect can be improved (1) by over-sizing the DM compared to the aperture ($\alpha$ parameter), (2) by using of a non-actuated reflective surface to extend the side length of the second DM, or (3) by using more wavelengths during the correction process, in the interaction matrix. However, as shown by Fig.~\ref{fig:oversizing_size}, the aperture extends exponentially with $F_0$ at small Fresnel numbers: it will eventually limits the contrast at small Fresnel numbers. In the setup used in this article ($\alpha$ = 10\%, and a second DM twice a large as the first one), we expect to be limited by this effect in contrast for $F_0\sim 5\times 10^2$. Finally, some DMs present very sharp features outside of the actuated region \citep[e.g. the Boston Micromachines --BMC-- DMs][]{mazoyer14b}, which can enhance this effect. We did not simulate those effects in this article. }

{All the effects shown in the vignetting section favor large Fresnel numbers, where they tend to disappear completely. Fig~\ref{fig:oversizing_size} and \ref{fig:oversizing_throughput} can be used to identified for which Fresnel number this effect will limit the performance. Note that the vignetting effects does not depends on which aperture is used (they are also present with a clear aperture). In the next Section, we introduce an opposite effect on the performance.}

%-------------------------------------------------------------------------------------------------
\subsection{Talbot effect (large Fresnel number effect)}
\label{sec:talbot}
%-------------------------------------------------------------------------------------------------

{The complete tools describing the behavior of the Talbot-effect-limited range ($F_0 \gg OWA^2$) are developed in Appendix~\ref{sec:annex}. In short, in this regime, the strokes necessary to correct for the discontinuity in the aperture increase linearly with the Fresnel number (see Eq. \ref{eq:sigma_DM2}). Eventually, for a finite number of actuators, the limiting contrast term is not the discontinuity in the aperture, but the frequency folding term \citep{giveon06,pueyo07} created by these important strokes. However, the amplitude of this term is directly related to the Fresnel number. Therefore, the larger is $F_0$, the more amplitude have to be corrected by the 2 DM system, even though the initial amplitude aberration in the aperture has not changed (see Eq.~\ref{eq:champDM2_withfrequencyfolder}). We showed in ACAD-OSM I that the increase of amplitude to be corrected impacts negatively both the contrast and throughput performance of the ACAD-OSM algorithm. This is the reason why, in the Talbot limited range $F_0 \gg OWA^2$, the performance both in contrast and in throughput are decreasing quickly with the Fresnel number. By choosing a favorable DM setup, not only the strokes necessary to correct for the aperture discontinuities are minimal, but the amount of amplitude to correct is also minimal. The actual position of the optimal Fresnel number for the Talbot effect is difficult to find analytically. In this section, we present a simple tool to put limits on the optimal Fresnel number.}

Starting from an initial periodic phase aberration, the optical path distance necessary for the phase to shift by one period (effectively reconstructing the original field) is known as the Talbot distance:
\begin{equation}
\label{eq:ztalbot}
z_{t}(N)  = \dfrac{2D^2}{\lambda_0 N^2}\,\,,
\end{equation}
where N is the number of cycles in the pupil in the original periodic map ($N \le N_{act}/2$) or the separation of a given speckle in focal plane. At half the Talbot distance, $z_{t}(N)/2$, the phase is the distance to have a $\pi$ phase shift. If the second DM is placed at that distance, it will only be able to introduce phase correction in the pupil plane (for this specific period). Therefore, if the distance $z$ between the DMs is exactly half the Talbot distance $z_{t}(N)/2$, a given speckle in the DH at a separation of N $\lambda_0/D_{ap}$ of the star can only be corrected in phase and not in amplitude.

On the other hand, a system with $z$ equal to a quarter of the Talbot distance $z_t(N)/4$ (the distance for the path to shift by a quarter wave) will allow the first DM to correct the phase, and the second to correct the amplitude only, for a given speckle in the DH at a separation of N $\lambda_0/D_{ap}$ of the star. This is the most effective distance to correct for phase and amplitude simultaneously. For this distance z, very small strokes on the DMs would ensure maximum effects at this specific separation N $\lambda_0/D_{ap}$.

To understand where to place the DMs for optimal performance in the DH, we invert the problem to show the ``worst" and ``best" corrected frequencies. For every given DM setup ($z$,$D$,$\lambda_0$),  the half-Talbot frequency $N_{t/2}$, is the position in the focal plane where this DM setup is completely ineffective at correcting amplitude aberrations, and the quarter-Talbot frequency $N_{t/4}$, is the position in the focal plane where this DM setup is the most effective for amplitude correction:

\begin{align}
\label{eq:freq_talbot}
N_{t/2}  = \dfrac{D}{\sqrt{\lambda_0 z}} = \sqrt{F_0}\,\,,\\
N_{t/4} = \dfrac{D}{\sqrt{2 \lambda_0 z}} = \sqrt{\dfrac{F_0}{2}} \,\,.
\end{align}

Assuming that the aberrations in the aperture have an decreasing power spectral density with N, the best cases for both contrast and throughput performance for this effect are when the quarter-Talbot frequency is within the DH ($IWA <N_{t/4}<OWA$), while keeping the half-Talbot frequency outside of the DH because in that case some modes of the DM start to be inefficient for amplitude correction ($N_{t/2}>OWA$). Therefore, in practice, the best Fresnel number depends on the chosen DH size. 

It is difficult to precisely identify the position of this minimum, but the use of these rules of thumbs can help put limits on its location: it is between when $IWA = N_{t/4}$ (if $N_{t/4}$ is smaller, the best frequency is outside of the DH) and when $N_{t/2} = OWA$ (the moment the non-corrected frequency enter the DH). The optimal Fresnel number $F_0$ for this effect is therefore between $2*IWA^2$ and $OWA^2$. 

For the contrast, several comparable terms are competing, but the degradation of the contrast for large $F_0$ follow a $F_0^{2}$ or $F_0^{4}$ law (see Eq. \ref{eq:Cres_apm} and \ref{eq:Cres_pha}). However, we show that whichever is the limiting contrast term, the bandwidth dependence always follows $(\lambda_0/\Delta \lambda)^2$, as already shown in ACAD-OSM I. Note that this contrast dependence with the BW is demonstrated in the Talbot limited case ($N_{t/2} \gg OWA$), and is not applicable elsewhere.

\subsection{Conclusion of the analytic formalism}
\label{sec:ccl_formalism}

In this section, we identified the major effects impacting contrast and throughput when changing the DM setup. The main points are:
\begin{itemize}
    \item All effects described here are only dependent on the Fresnel number and not of $D$, $z$ or $\lambda_0$ independently. They are not dependent of the coronagraph nor on the aperture.
    \item Because the vignetting effect, the contrast and the throughput improves with $F_0$ at small Fresnel numbers. The effect on contrast can be corrected for in monochromatic light, but not for large BWs. This effect can probably be minimized by expanding greatly the reflective area around the second DM.
    \item Because of Talbot effect, the performance in throughput and contrast degrades with $F_0$ increasing. The contrast degrades with the Fresnel number at large $F_0$, as $F_0^2$ or $F_0^4$ (see Appendix~\ref{sec:annex}). This effect is inherent to contrast correction on a finite DH.
\end{itemize}

These 3 points clearly show that for each coronagraph, aperture, or method of correction, there is an optimal DM setup (or a few optimal DM setups) for best contrast and throughput performance, only dependent of the Fresnel number $F_0$. For too large or too small Fresnel numbers, the performance in contrast and throughput inevitably degrades. For a given aperture and a given DH geometry detailed simulations have to be run to find this optimal DM setup. We run these simulations for the aperture of WFIRST in the next section.

%-------------------------------------------------------------------------------------------------
%-------------------------------------------------------------------------------------------------
%-------------------------------------------------------------------------------------------------
\section{Impact of DM setup: complete numerical simulations}
\label{sec:DM_setup}
%-------------------------------------------------------------------------------------------------
%-------------------------------------------------------------------------------------------------
%-------------------------------------------------------------------------------------------------

\subsection{Numerical simulation parameters}
\label{sec:DM_setup_parameter}

In this section, we study the influence of the DM setup on the performance in terms of contrast level, throughput, and resistance to small TT jitter, for a WFIRST-like aperture (Figure~\ref{fig:wfirst_ls46_optimdm_dh}, top, left). Contrary to more friendly apertures, this difficult aperture (36\% central obscuration, large struts covering 9\% of the left area) offer a wide range of performance with the ACAD-OSM technique (contrast levels from $10^{-5}$ to $10^{-12}$ and throughput from a few percents to several tens of percents), which is ideal for comparing the impact of DM setup on these parameters. The same type of coronagraphs as the ones described in Sec.~\ref{sec:scda} are used: an APLC, a charge 6 PAVC and a charge 4 RAVC. The parameters of these coronagraphs are summarized in Table~\ref{tab:ularasa}. We do not study the size of the DH because it has been studied intensively in simulation \citep{borde06, beaulieu17} and experimentally \citep{mazoyer13c}. The DH is fixed to 3-10 $\lambda_0/D_{ap}$ for the vortex coronagraphs and 5-12 $\lambda_0/D_{ap}$ for the APLC. The wavelength is 550 nm and the BW 10$\%$. 

As explained in Section~\ref{sec:vignetting_throughput}, the second mirror is extended to up to twice the side-length of the DM by surrounding the DM with a non-actuated reflective surface. The DMs are oversized compared to the pupil by $\alpha = 10\%$. The number of interaction matrices used for these corrections is usually 8. However, for some difficult cases (\textit{i.e.} those requiring the highest strokes, with large Fresnel number), the performance in contrast could still improve when increasing the number of interaction matrices used. This shows that the local contrast minimum was not obtained after 8 matrices. Because the ACAD-OSM algorithm is contrast performance driven, we tried, in this section, to always reach the local minimum in contrast, by increasing the number of matrices above 8 when needed, until no improvement of contrast is observed.

%-------------------------------------------------------------------------------------------------
\begin{table*}
\centering
\caption{Half-Talbot frequency $N_{t/2}$ and Fresnel number $F_0$ for several DM setups. The wavelength is always 550 nm, the number of actuators 48.}
\label{tab:zlsurD_Nopt}
\begin{tabular}{|c|c|c|c|c|}
\hline
    \diagbox{D}{z}        & 0.3 m                                                                                             & 0.7 m                                                                                             & 1.0 m                                                                                               & 1.5 m                                                                                             \\ \hline
48 * 0.1 mm & \begin{tabular}[c]{@{}c@{}}$F_0= 1.4 \times 10^{2}$\\ $N_{t/2} = 12 \lambda/D$\end{tabular}  & \begin{tabular}[c]{@{}c@{}}$F_0= 6.0 \times 10^{1}$\\ $N_{t/2} = 7.7 \lambda/D$\end{tabular} & \begin{tabular}[c]{@{}c@{}}$F_0= 4.2 \times 10^{1}$\\ $N_{t/2} = 6.5 \lambda/D$\end{tabular} & \begin{tabular}[c]{@{}c@{}}$F_0= 2.8 \times 10^{1}$\\ $N_{t/2} = 5.3 \lambda/D$\end{tabular} \\ \hline
48 * 0.3 mm & \begin{tabular}[c]{@{}c@{}}$F_0= 1.3 \times 10^{3}$\\ $N_{t/2} = 35 \lambda/D$\end{tabular}  & \begin{tabular}[c]{@{}c@{}}$F_0= 5.4 \times 10^{2}$\\ $N_{t/2} = 23 \lambda/D$\end{tabular}  & \begin{tabular}[c]{@{}c@{}}$F_0= 3.8 \times 10^{2}$\\ $N_{t/2} = 19 \lambda/D$\end{tabular}  & \begin{tabular}[c]{@{}c@{}}$F_0= 2.5 \times 10^{2}$\\ $N_{t/2} = 16 \lambda/D$\end{tabular}    \\ \hline
48 * 0.7 mm & \begin{tabular}[c]{@{}c@{}}$F_0= 6.9 \times 10^{3}$\\ $N_{t/2} = 83 \lambda/D$\end{tabular}  & \begin{tabular}[c]{@{}c@{}}$F_0= 2.9 \times 10^{3}$\\ $N_{t/2} = 54 \lambda/D$\end{tabular}  & \begin{tabular}[c]{@{}c@{}}$F_0= 2.1 \times 10^{3}$\\ $N_{t/2} = 45 \lambda/D$\end{tabular}  & \begin{tabular}[c]{@{}c@{}}$F_0= 71.4 \times 10^{3}$\\ $N_{t/2} = 37 \lambda/D$\end{tabular}  \\ \hline
48 * 1.0 mm   & \begin{tabular}[c]{@{}c@{}}$F_0= 1.4 \times 10^{4}$\\ $N_{t/2} = 118 \lambda/D$\end{tabular} & \begin{tabular}[c]{@{}c@{}}$F_0= 6.0 \times 10^{3}$\\ $N_{t/2} = 77 \lambda/D$\end{tabular}  & \begin{tabular}[c]{@{}c@{}}$F_0= 4.2 \times 10^{3}$\\ $N_{t/2} = 64 \lambda/D$\end{tabular}  & \begin{tabular}[c]{@{}c@{}}$F_0= 2.8 \times 10^{3}$\\ $N_{t/2} = 52 \lambda/D$\end{tabular}  \\ \hline
\end{tabular}
\end{table*}
%-------------------------------------------------------------------------------------------------

%%% ---------------------------------------------------------------------------
\begin{figure}
\begin{center}
 \includegraphics[ trim= 1.5cm 0.9cm 1.0cm 0.5cm, clip = true, width = .48\textwidth]{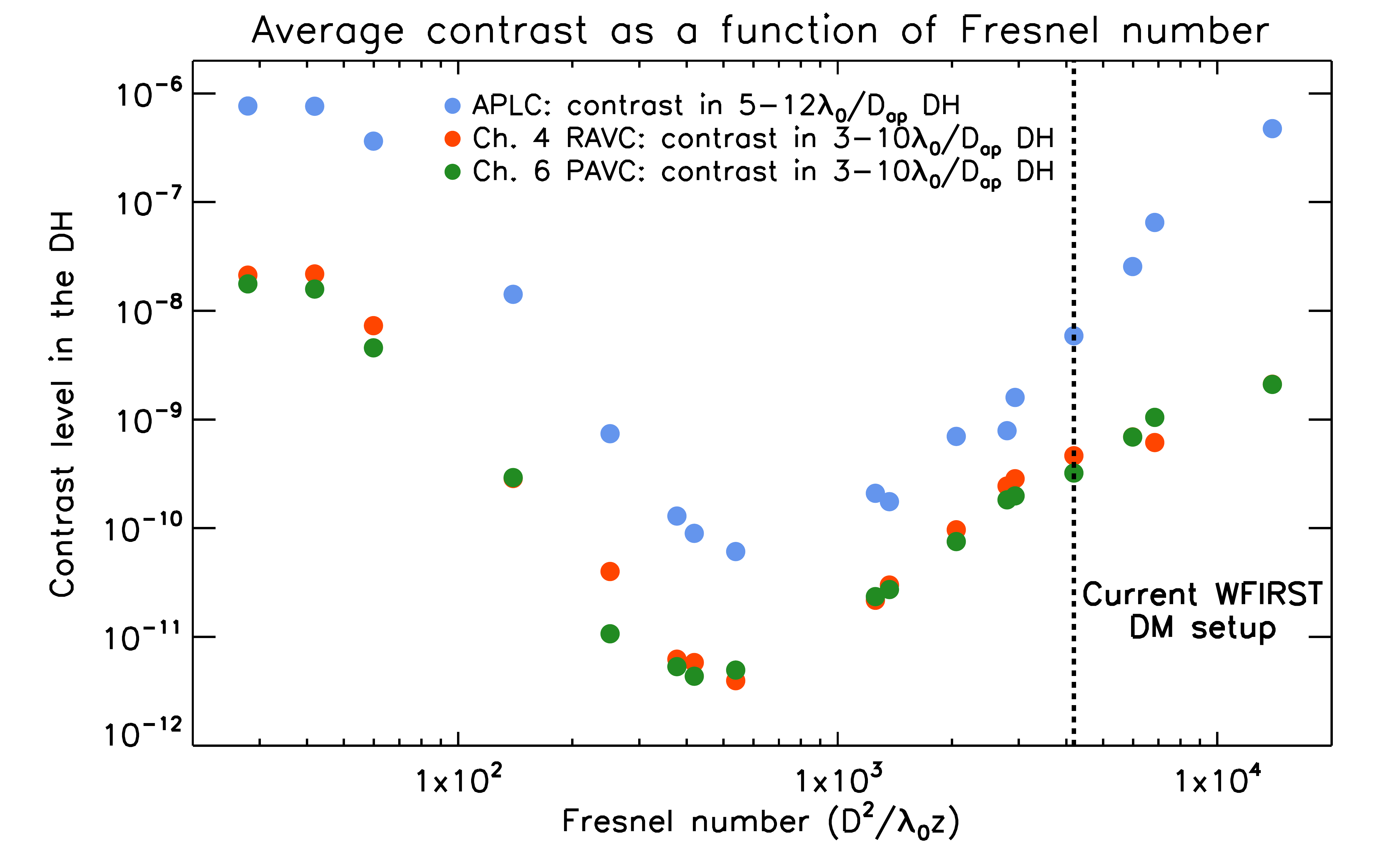}
 \end{center}
\caption[plop]
{\label{fig:perf_contrast_fresnel} Performance of the correction in contrast as a function of the Fresnel number, for the WFIRST aperture, a 10\% BW around 550 nm, and for 48 actuators. The vertical black dashed line indicate the actual WFIRST DM setup.}
\end{figure}
%%% ---------------------------------------------------------------------------

To test the DM setup with a fixed number of actuators, we change the IAP. We set $N_{act}$ = 48 actuators and choose four values for the IAP: 0.1 mm (\textit{i.e.} 0.48 cm DMs), 0.3 mm (which is the IAP of the BMC DMs and corresponds in that case to 1.44 cm DMs), 0.7 mm (\textit{i.e.} 3.3 cm DMs), and 1 mm (\textit{i.e.} 4.8 cm DMs). This last case (48 actuators with a 1mm IAP) corresponds to the Xinetics DMs selected for the WFIRST mission. For each of the cases studied, we measured the Fresnel number $F_0$ (Eq.~\ref{eq:fnum}) and the half-Talbot frequency $N_{t/2}$ (Eq.~\ref{eq:freq_talbot}), shown in Table~\ref{tab:zlsurD_Nopt}. We also tested the case where IAP = 0.1, $D = 48 * 0.1$ mm and $z$ = 0.1 m ($F_0= 4.2 \times 10^{2}$, $N_{t/2} = 20 \lambda_0/D_{ap}$), to show that the degradation of performance we observed with the smallest DM size (IAP = 0.1 mm, $D = 48 * 0.1$ mm, first line in Table~\ref{tab:zlsurD_Nopt}) fits in the general Fresnel number trend and is not specifically due to the small size of the DM.

Fig. \ref{fig:perf_contrast_fresnel} shows the performance in contrast levels and Fig. \ref{fig:perf_throughput_fresnel} shows the performance in throughput, as a function of the Fresnel number for the three coronagraphs. The first interesting aspect is that points with different distances and DM sizes but with similar Fresnel numbers gives similar performance in contrast and throughput, proving that the performance is not dependent on $z$ or $D$ independently but only on the Fresnel number. For example, DM setups with IAP = 1 mm, $D = 48*1$ mm and $z$ = 1.5 m ($F_0= 2.8 \times 10^{3}$) and with IAP = 0.7 mm, $D = 48*0.7$ mm and $z$ = 0.7 m ($F_0= 2.9 \times 10^{3}$) obtain similar results. The vertical black dashed line indicates the actual WFIRST DM setup (IAP = 1 mm, $D = 48*1$ mm and $z$ = 1 m). 

\subsection{Contrast performance as a function of Fresnel number}
\label{sec:contrast_fresnel}

The performance in contrast (Fig.~\ref{fig:perf_contrast_fresnel}) increases and then decreases after a sweet spot located around $F_0 = 5 \times 10^{2}$. \\

\subsubsection{Behavior at low Fresnel number}
{As expected, for small Fresnel numbers, the contrast increases with $F_0$. As shown in Fig.~\ref{fig:contrast_vignetting}, the correction process with 3 discrete wavelengths actually goes deep, but when applied to a more continuous BW to measure the performance, these DM shapes provide a contrast that worsens as the Fresnel number decrease. This means that the performance is good for monochromatic light (or even for 3 discrete and separated wavelengths) but not for a continuous BW. We showed in Sec.~\ref{sec:vignetting_contrast} that this effect is limiting the correction only for  $F_0<5 \times 10^{2}$. 

An increase in the number of wavelengths concatenated in the interaction matrix or an extension of the reflective non actuated zone outside of the second DM would probably shift the optimal Fresnel number towards smaller Fresnel numbers. These solutions are computationally heavier and were not tried in this paper. A line fitted to the slope in that regime showed that contrast degrades as $F_0^4$.}

\subsubsection{Behavior at large Fresnel number}

For $F_0>5\times10^{2}$, we obtain the same performance in contrast for 3 discrete wavelengths and for a more continuous BW, which shows that the contrast is not limited by the vignetting of the Gibbs ring. We are in the Talbot effect-limited regime. We fit a line to the increasing slope of this curve and found contrast degrades as $F_0^{2}$ as expected from Appendix~\ref{sec:annex}.

%%% ---------------------------------------------------------------------------
\begin{figure}
\begin{center}
 \includegraphics[trim= 2.5cm 0.9cm 1.0cm 0.5cm, clip = true, width = .48\textwidth]{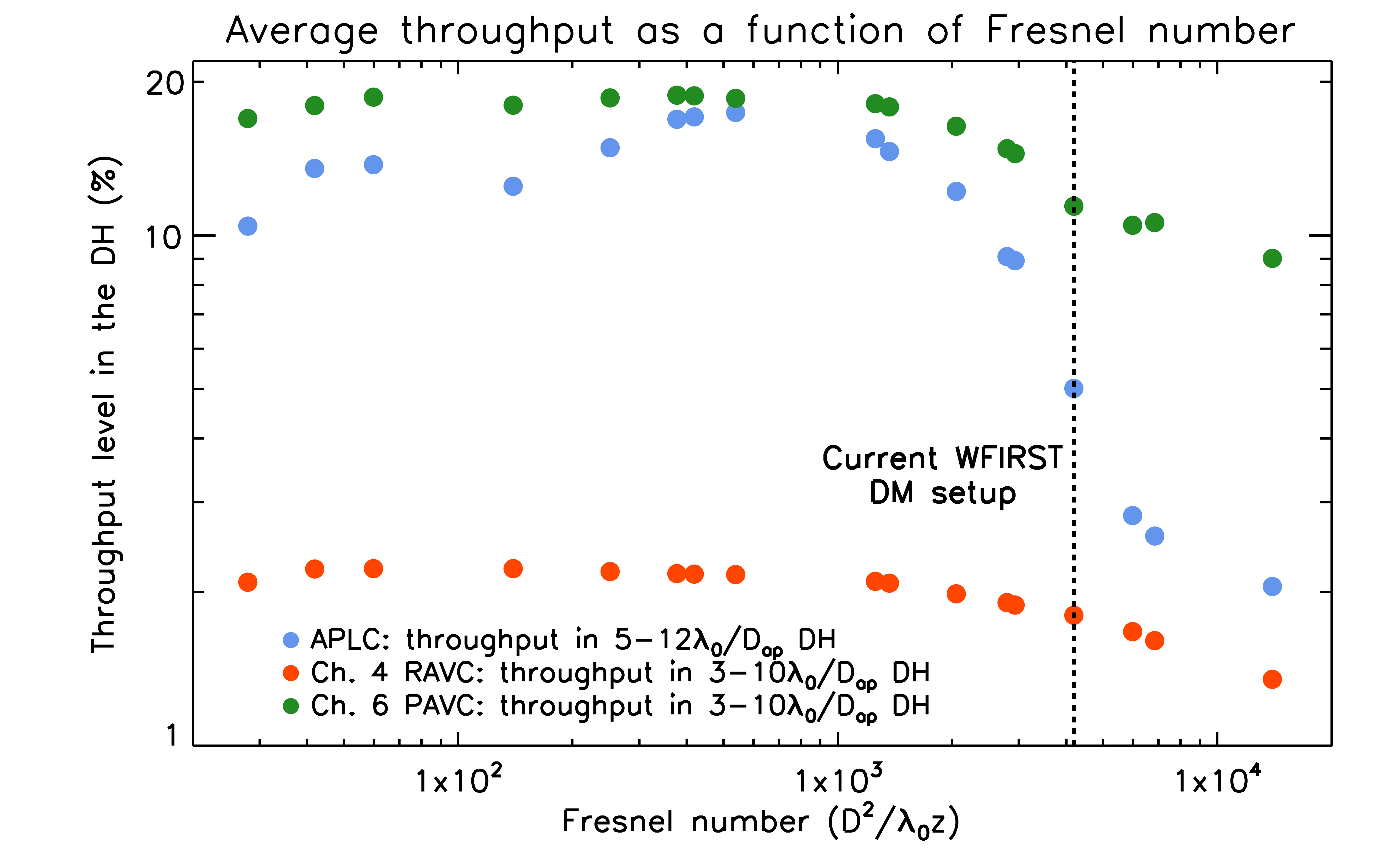}
 \end{center}
\caption[plop]
{\label{fig:perf_throughput_fresnel} Performance of throughput in contrast as a function of the Fresnel number for the WFIRST aperture, a 10\% BW around 550 nm, and for 48 actuators. The vertical black dashed line indicates the actual WFIRST DM setup.}
\end{figure}
%%% ---------------------------------------------------------------------------

\subsubsection{Optimum Fresnel number}

{The optimal Fresnel number observed in this case is $F_0\sim 5 \times 10^{2}$. 

We show in Section~\ref{sec:talbot} that the minimum due to the Talbot propagation is located for $F_0$ between $2 *IWA^2$ and $OWA^2$ which corresponds to 18-100 for a 3-10 $\lambda_0/D_{ap}$ DH and 50-144 for a 5-12 $\lambda_0/D_{ap}$.

In Section~\ref{sec:vignetting_throughput}, we showed that the effect of vignetting starts when the diffracted aperture in the second DM plane reaches the edge of the second DM. In the current DM setup (the DMs are 10\% larger than the aperture, and we surrounded the the second DM with a non-actuated reflective surface up to twice the size of the DM), this corresponds to $F_0= 5\times 10^2$ (Fig~\ref{fig:oversizing_size}).

This means that we are currently limited by the vignetting effect and not by the Talbot effect. With larger optics, the optimal Fresnel number could probably be shifted to the left. Further development are necessary to precisely locate the optimum Fresnel point as a function of the correction parameters. 

However, we show in this section that the type of coronagraph has no influence on the position of the sweet spot. We do not know if this optimal range of Fresnel numbers is dependent on the ACAD-like technique used \citep[e.g.][]{krist16}. However, because the effects described in the previous theoretical section are independent of the technique used, we know that every 2 DM techniques for coronagraphy correction will present an optimum point or region for contrast and throughput. }

%-------------------------------------------------------------------------------------------------
\subsection{Throughput performance as a function of Fresnel number}
\label{sec:throughput_fresnel}
%-------------------------------------------------------------------------------------------------

Fig.~\ref{fig:perf_throughput_fresnel} shows the variation of throughput performance. The vignetting of the chromatic Gibbs rings is the limiting effect at small Fresnel numbers. We chose in this section to surround the second DM with a non-actuated reflective surface up to twice the size of the DM.  
We showed in Section~\ref{sec:vignetting_throughput} that, without this assumption, we would have a important loss of off-axis throughput for $F_0< 2\times10^{2}$ due to the 2 DM system vignetting. However, the Talbot effect, described in Section~\ref{sec:talbot} and Appendix~\ref{sec:annex} (limiting effect at large Fresnel number) degrades the throughput with $F_0$. Witth this assumption the throughput performance is almost flat until $F_0\sim 10^{3}$ but then decreases quickly at higher Fresnel numbers.

The main point of this section is that the DM setup, as opposed to to the parameters described in Sec.~\ref{sec:fine_tunning} can be optimized to maximize both the contrast and throughput performance at the same time. Indeed $F_0= 5 \times 10^{2}$ corresponds simultaneously to a maximum in contrast and in throughput performance.

%-------------------------------------------------------------------------------------------------
\subsection{Robustness to small tip-tilt jitter as a function of Fresnel number}
\label{sec:jitter_fresnel}
%-------------------------------------------------------------------------------------------------

\begin{figure}
\begin{center}
 \includegraphics[trim= 1.5cm 0.9cm 1.0cm 0.5cm, clip = true, width = .48\textwidth]{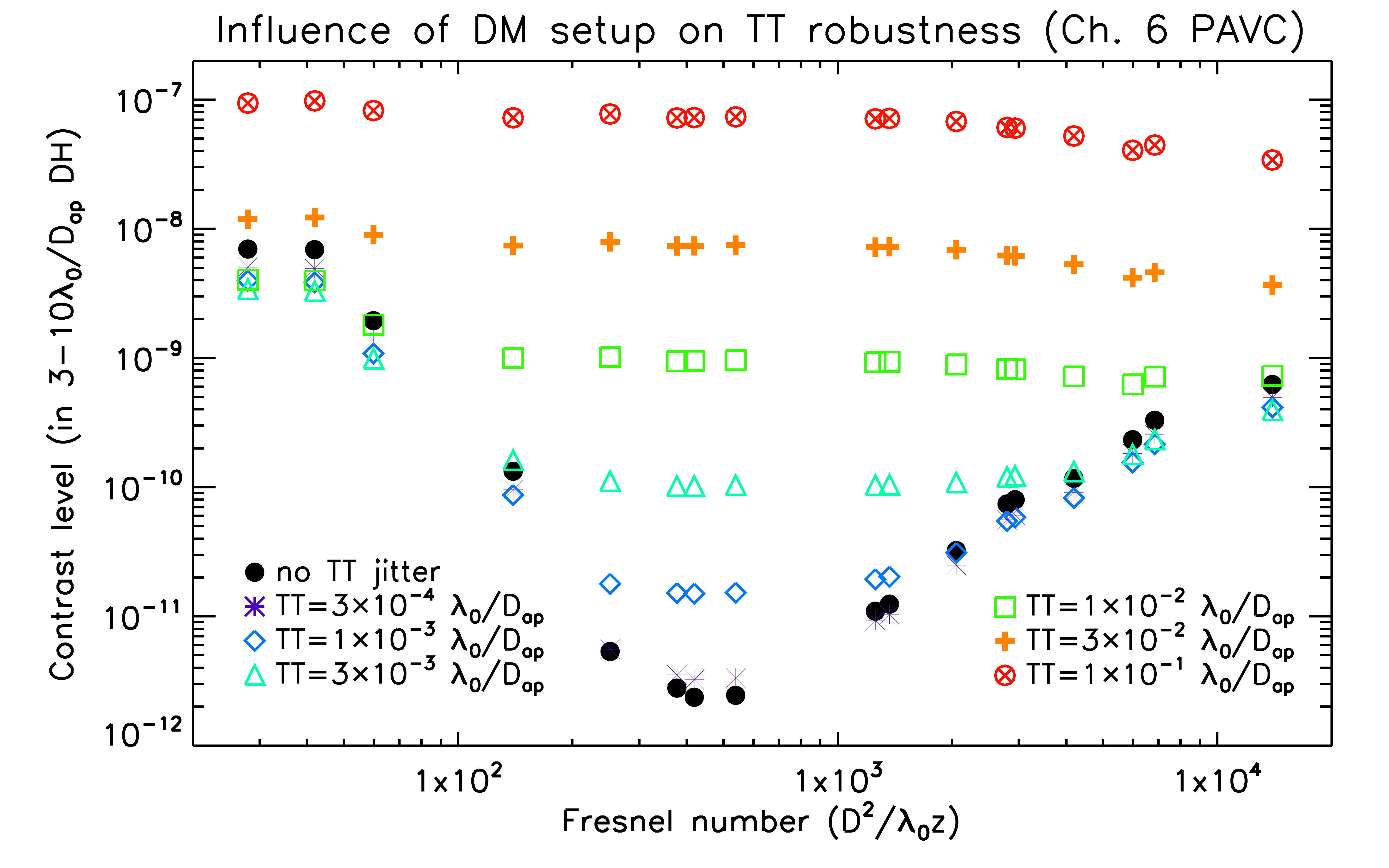}
 \includegraphics[trim= 1.5cm 0.9cm 1.0cm 0.5cm, clip = true, width = .48\textwidth]{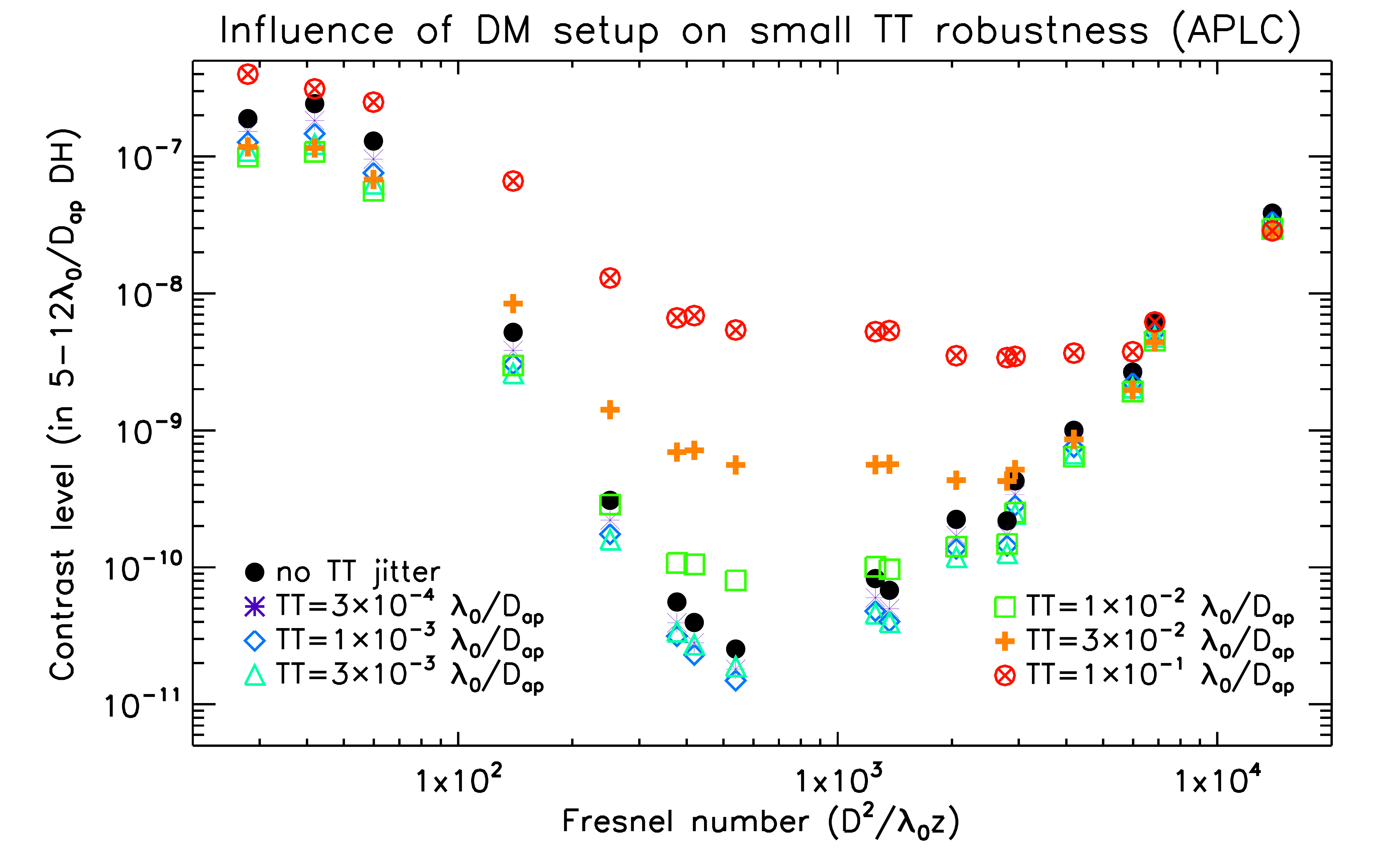}
 \end{center}
\caption[plop]
{\label{fig:perf_f_F_jitter} Influence of the Fresnel number $F_0$ on small TT jitter robustness for a charge 6 PAVC (top) and an APLC (bottom). These were obtained with a WFIRST aperture, a 10\% BW around 550 nm, and for 48 actuators.}
\end{figure}

Finally, Figure~\ref{fig:perf_f_F_jitter} shows the influence of the Fresnel number on small TT jitter robustness for two coronagraphs (charge 6 PAVC and APLC). The degradation of contrast level in the DH for different levels of TT jitter introduced is shown with different symbols (from $3 \times 10^{-4}$ to $1 \times 10^{-1} \lambda_0/D_{ap}$). These two plots show that the TT jitter robustness is mainly dependant on the type of coronagraph and not on the DM setup.

For example, for the charge 6 coronagraph, and for a TT jitter of $3 \times 10^{-3} \lambda_0/D_{ap}$ (blue triangles), the design is insensitive to this jitter when the performance of the system are worse than $1 \times 10^{-10}$ in contrast (i.e. for $F_0< 10^{2}$ and $F_0> 3 \times 10^{3}$). However, when the performance in contrast of the system ais better than $10^{-10}$ (i.e. for $10^{2} < F_0<3 \times 10^{3} $), the design is limited by the TT jitter to this level and the performance scarcely depends on the DM setup. Using an higher TT level $1 \times 10^{-1} \lambda_0/D_{ap}$ (red circles with Xs), the contrast level is always limited by the TT and not by the performance of the coronagraph and in that case is mostly constant (variation of a factor of 3 only on for the whole $F_0$ range studied). We observe the same results on the APLC, except that this coronagraph presents a greater robustness to contrast which was to be expected.

%%% ---------------------------------------------------------------------------
%%% ---------------------------------------------------------------------------

These results also show that for large primary mirror instruments, where the diameter of the star can be resolved, the effects of the DM setup on contrast performance will often be negligible compared to the effects of the star size. This means that a large range of DM setups will have the same contrast performance because of TT jitter or star diameter, mainly driven by the chosen coronagraphic design. However, this is not true for the performance on off-axis throughput, where the optimal DM setup should always be preferred.

%-------------------------------------------------------------------------------------------------
\subsection{Influence of central wavelength}
\label{sec:wavelength}
%-------------------------------------------------------------------------------------------------

The analysis of the performance as a function of the central wavelength $\lambda_0$ easily fits in the Fresnel propagation model we defined in the previous section. For the same DM setup, increasing the central wavelength will reduce the Fresnel number and enhance the vignetting effect, but decreasing the central wavelength will increase the Fresnel number and the correction is more likely to be limited by the Talbot effect. 

For small changes in the central wavelength, this can be compensated by changing only the distance between the DMs (e.g. is you increase the central wavelength by $10\%$, the same performance can be obtained by reducing the distance $z$ by 10\%). Note that this is important for instruments that will try to achieve high performance both in the visible and in the IR: the sweet spot cannot be achieved for two very different wavelengths with the same DM setup.

However, one has to be careful when extending these results to all wavelengths because it has been developed using a Fresnel assumption. A given DM stroke can be simulated with the Fresnel assumption at some wavelengths but may not at a shorter wavelength. \\

In this section, we showed that the DM setup (D,z) only impacts the performance of the correction through the Fresnel number. We found a sweet spot corresponding both to a maximum in contrast and in throughput. We also showed that the robustness to TT jitter is mostly insensitive to the DM setup. We show that these results are not dependent on the coronagraph. A significant consequence of this section is that the DM setup currently chosen for the WFIRST mission is probably not optimal for either contrast or throughput performance. In the next section, we show how we can optimize ACAD-OSM for future missions, using the WFIRST mission as an example. 

%-------------------------------------------------------------------------------------------------
%-------------------------------------------------------------------------------------------------
%-------------------------------------------------------------------------------------------------
\section{Optimization for future missions}
\label{sec:optim_missions}
%-------------------------------------------------------------------------------------------------
%-------------------------------------------------------------------------------------------------
%-------------------------------------------------------------------------------------------------

%-------------------------------------------------------------------------------------------------
\subsection{Optimization of the ACAD-OSM technique: case of the WFIRST aperture}
\label{sec:optim_wfirst}
%-------------------------------------------------------------------------------------------------
In the previous section, we showed that we could reach performance in contrast as good as $5\times10^{-12}$ (mean contrast in the DH) with the WFIRST aperture and an optimal DM setup. However, as explained in Section~\ref{sec:fine_tunning}, contrast is not the only metric to consider to optimize the detection of exoplanets. Also, we showed in Section~\ref{sec:jitter_fresnel} that the performance in contrast with a point-like star in the absence of jitter does not represent the final detection limit of a given instrument. For this reason, in this section we provide an example of how to optimize the ACAD-OSM technique in a realistic case, with a different approach. For the WFIRST aperture and the charge 6 PAVC, we will set the contrast goal to $5\times10^{-10}$ ($\sim 10^{-9.3}$), which is the contrast limit set by a $1\times10^{-2}$ jitter (green squares in Fig~\ref{fig:perf_f_F_jitter}, top). For this contrast goal, we show that with a careful optimization of the DM setup, ACAD-OSM operating point and size of the Lyot stop radius, we can double the performance in throughput inside the DH, for the same contrast performance.

\begin{figure}
\begin{center}
 \includegraphics[width = .48\textwidth, trim= 1cm 0.8cm 0.7cm 0.3cm, clip = true]{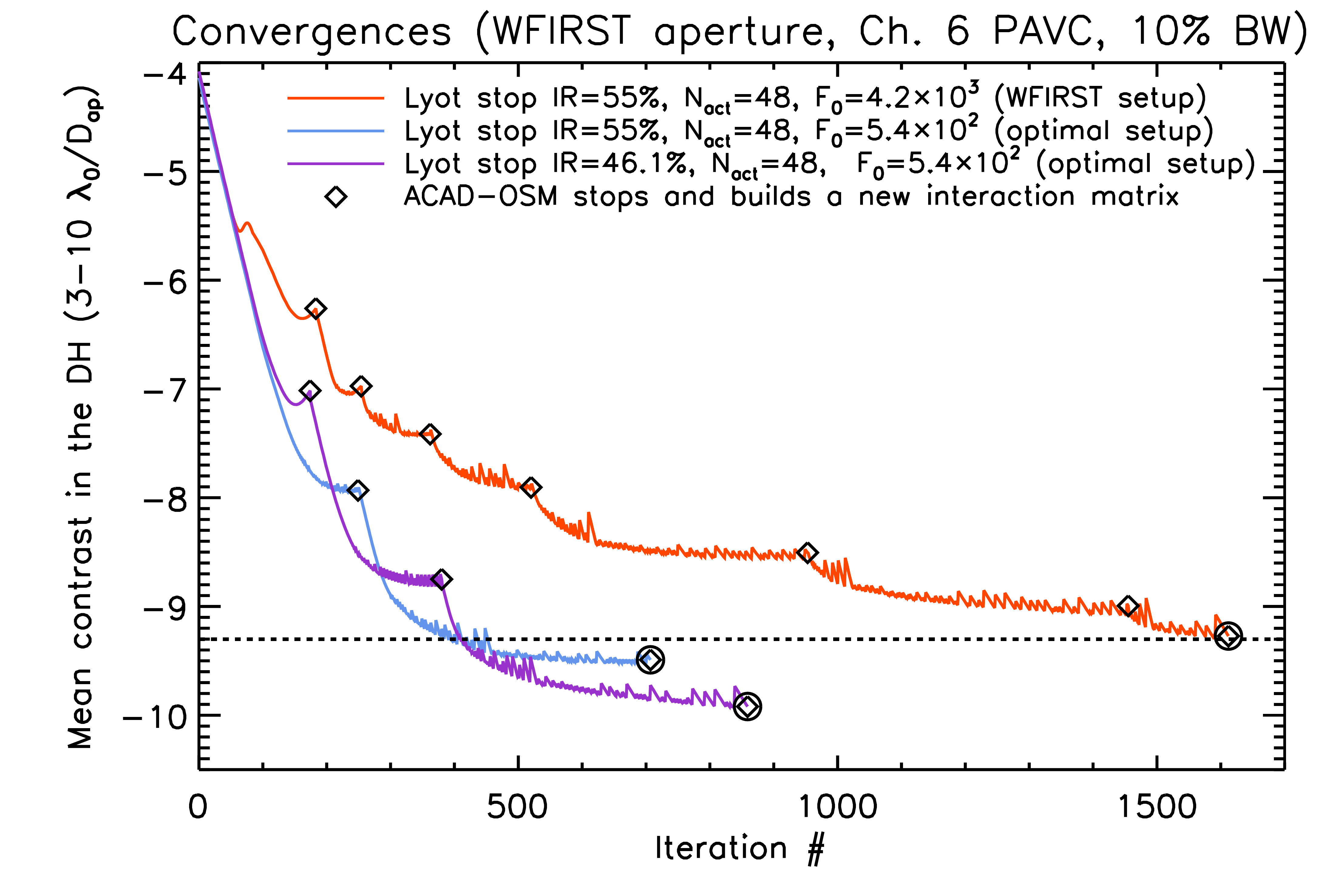}
 \end{center}
\caption[]
{\label{fig:iter_dmsetup} Convergence of the mean contrast level in the DH as a function of the number of iterations for the ACAD-OSM solution. The black circles show the DM shapes selected when they achieve at least the contrast goal.}
\end{figure}

Fig.~\ref{fig:iter_dmsetup} shows the mean contrast level as a function of the iteration number for several DM setups and coronagraphs. All use the WFIRST aperture, the same number of actuators ($48\times48$ actuators) and have the same BW ($\Delta \lambda /\lambda_0 = $ 10\%, centered around 550 nm) but with different DM setups and charge 6 PAVCs. The black diamonds show the iterations at which the ACAD-OSM algorithm stops and builds a new interaction matrix. The contrast goal of $10^{-9.3}$ is represented by a black, dashed line. For each of these configuration, we show the performance in throughput in Fig~\ref{fig:wfirst_find_optim_throughput}.

We start by the configuration used as an example in paper ACAD-OSM I (Sec. 2, Fig. 5), with a charge 6 PAVC with a Lyot Stop inner radius of 55\% and the WFIRST DM setup ($F_0= 4.2\times 10^{3}$). The contrast at each iteration for this configuration is shown in red in Fig.~\ref{fig:iter_dmsetup}. We select for operating point the result obtained after 7 matrices only, represented by the circled black diamond. By not letting the correction to reach the local minimum ($10^{-9.5}$, shown in red in Fig~\ref{fig:sizeLS_throughput}), we ensure that we obtain a slightly better throughput (see Section~\ref{sec:Matrix_number}). We plot the result in throughput in Fig~\ref{fig:wfirst_find_optim_throughput} (red curve). 

This DM setup used here is not optimal. The high strokes necessary to correct for the aperture discontinuities increase the amplitude to correct for, which limits the performance in throughput and contrast. For this reason, the DM setup found to be optimal in the previous section is used: $F_0= 5.4\times 10^{2}$. For this DM setup and the same charge 6 PAVC with a Lyot Stop inner radius of 55\%, the $10^{-9.3}$ goal is achieved in 2 iterations only. We also plot the throughput on Fig~\ref{fig:wfirst_find_optim_throughput} (blue curve). The less important DM strokes allow us to achieve a more important throughput. However, whatever is the DM setup, the performance will be limited by the throughput of the coronagraph itself represented by a black dot-dashed line).

\begin{figure}
\begin{center}
 \includegraphics[width = .48\textwidth, trim= 0.1cm 4.5cm 5cm 4cm, clip = true]{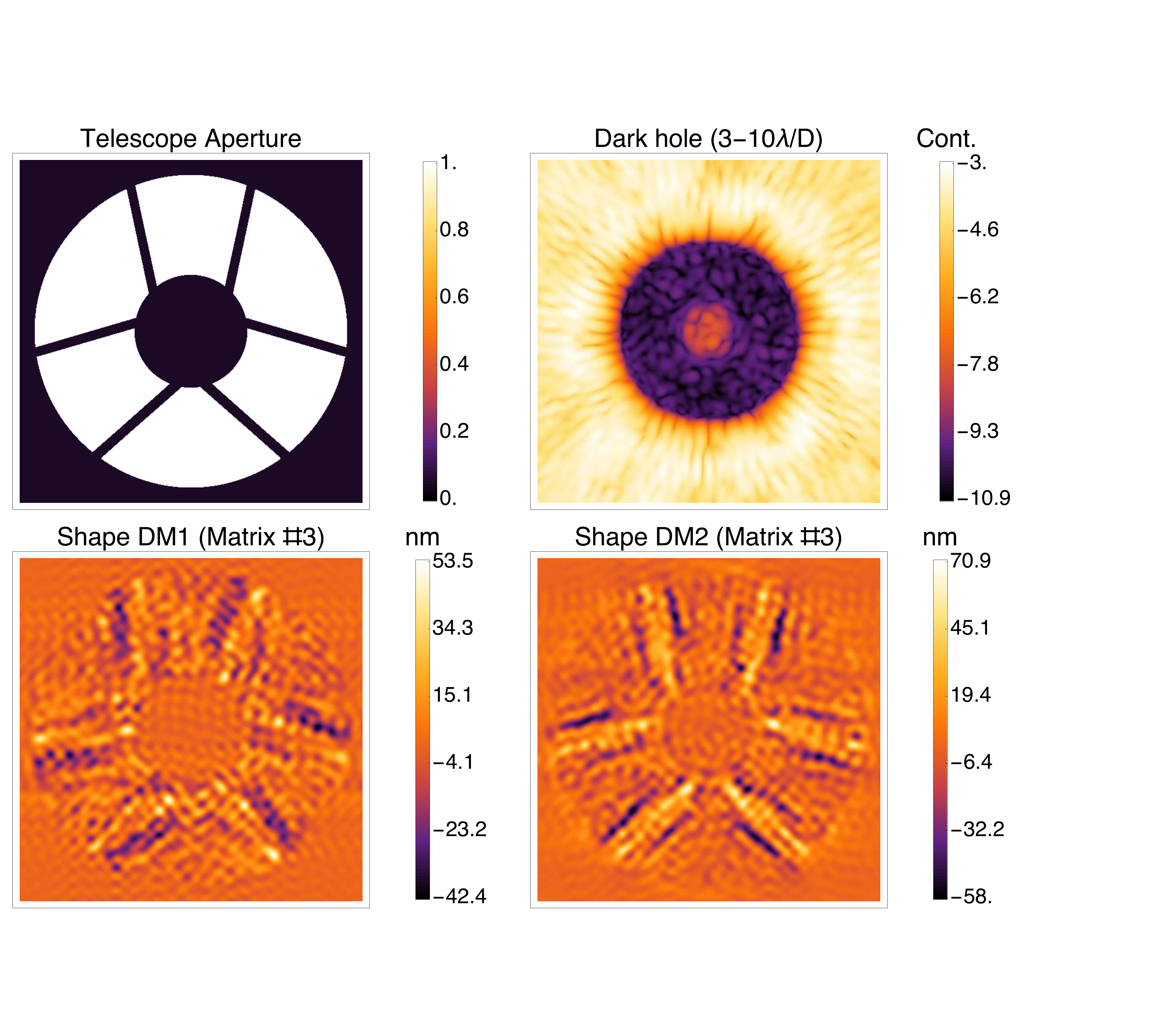}
 \end{center}
\caption[]
{\label{fig:wfirst_ls46_optimdm_dh} Results for the WFIRST aperture a charge 6 PAVC coronagraph with a 46.1\% Lyot stop radius, $N_{act} = 48$, IAP = 0.3 mm, D = $48 * 0.3$ mm, $z = 0.3$ m and $\Delta \lambda /\lambda_0 = $ 10\% BW). Top left: WFIRST aperture. Top right: the final DH. Bottom: the DM shapes obtained using ACAD-OSM for this solution obtained after the third matrix.}
\end{figure}

\begin{figure}
\begin{center}
 \includegraphics[trim= 1.5cm 0cm 0cm 0cm, clip = true,width = .48\textwidth]{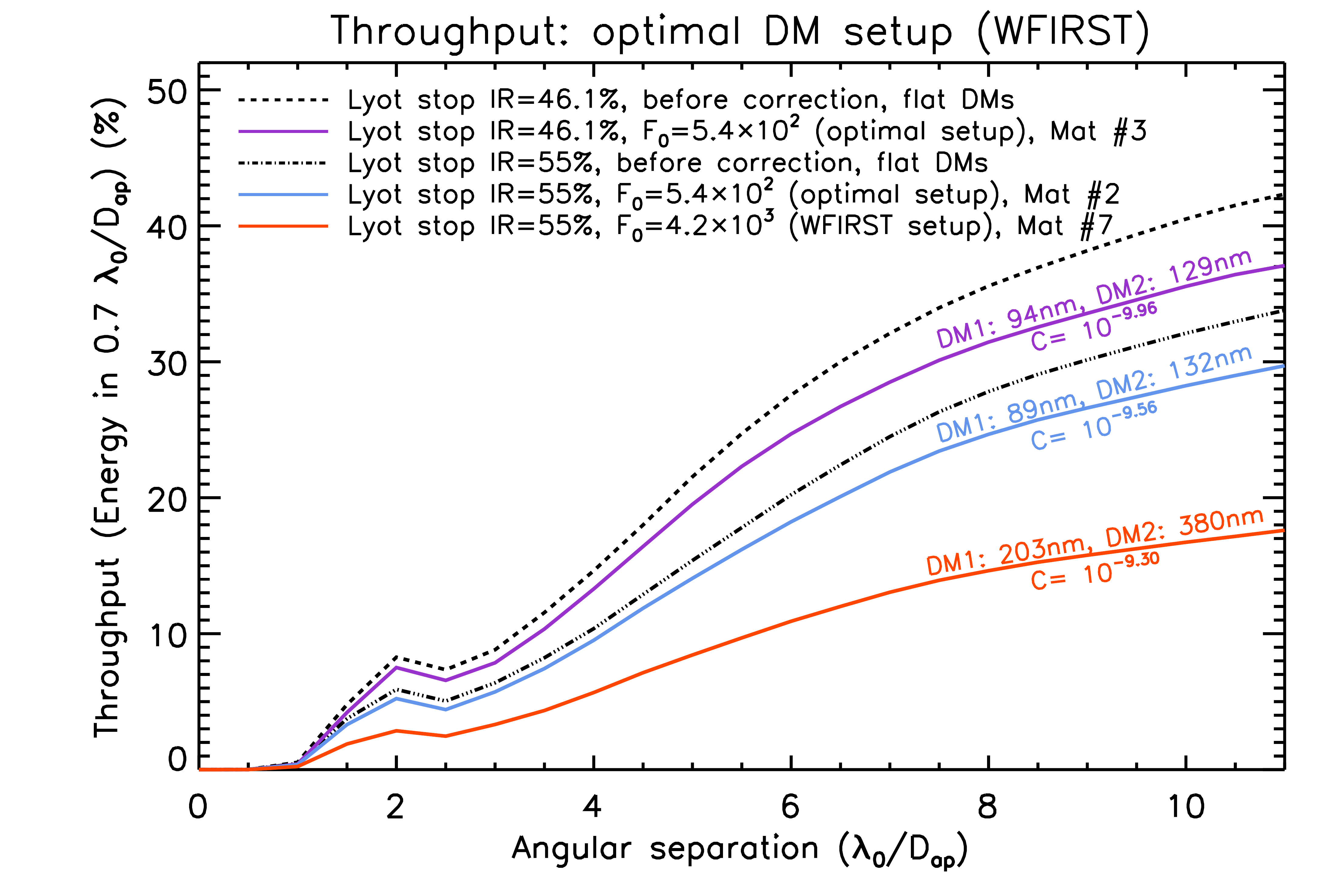}
  \end{center}
\caption[]
{\label{fig:wfirst_find_optim_throughput} Throughput results for different DM set-ups and coronagraphs, with the WFIRST aperture ($\Delta \lambda /\lambda_0 = $ 10\% BW). The dashed and dot-dashed line shows the throughput with flat DMs, before any correction (due to the PAVCs alone). The solid lines show the throughput after ACAD-OSM corrections. We also show the contrast and strokes (PV) on the same plot.}
\end{figure}

For this reason, and because the local minimum of this configuration is far better than the contrast goal we have set (Fig.\ref{fig:perf_contrast_fresnel} shows that we can achieve $10^{-11.3}$ with this configuration), we try to change the coronagraph with a smaller inner radius Lyot stop, which will degrade the contrast but improve the throughput (Section~\ref{sec:LSradius}). A charge 6 PAVC with a 46.1\% inner radius of the Lyot stop is used here. In Fig.~\ref{fig:iter_dmsetup}, purple line, we show the result obtained with this setup. We notice that because the Lyot stop inner radius is smaller, the contrast increase due to each matrix is less important compared to the larger Lyot case (red curve). For this DM setup ($F_0= 5.4\times 10^{2}$) and the charge 6 PAVC with a Lyot Stop inner radius of 46.1\%, the operating point at $10^{-9.3}$ contrast is achieved in 3 interaction matrices. We select these DM shapes and draw the throughput curve in Fig.~\ref{fig:wfirst_find_optim_throughput} (purple line). We show that using this setup, we can obtain a $10^{-9.96}$ contrast with a throughput ranging from 8\% to 37\% in the 3-10 $\lambda_0/D_{ap}$ DH. {As in the case of the SCDA aperture (Sec.~\ref{sec:scda}), the loss of throughput due to ACAD-OSM only (difference between the black dashed and purple lines) is only of a few percents: the final throughput level is mainly driven by the coronagraph design only.} We show the final DH and DM shapes obtained in this case in Fig.~\ref{fig:wfirst_ls46_optimdm_dh}. 

We showed that with a careful optimization of the coronagraph and of the DM setup, we can decrease the strokes level by a factor of 3 and improve the performance in throughput by more than a factor 2 inside the DH. We end this section by discussing the possibility of optimization on real ground- or space-based missions.

%-------------------------------------------------------------------------------------------------
%-------------------------------------------------------------------------------------------------

%-------------------------------------------------------------------------------------------------
%-------------------------------------------------------------------------------------------------

%-------------------------------------------------------------------------------------------------
%-------------------------------------------------------------------------------------------------

%-------------------------------------------------------------------------------------------------
\subsection{Discussion of future high-contrast instruments}
\label{sec:Discutions}
%-------------------------------------------------------------------------------------------------

We showed in the previous section how to optimize the ACAD-OSM technique by changing the size of the Lyot stop and the DM setup to optimize the performance for a given contrast goal. This optimization has to be done for each coronagraph, aperture, and size of the DH to optimize ACAD-OSM. 

However, when presenting the results of this paper, we purposely did not take into account several aspects of the feasibility of the system we studied. The range of Fresnel numbers we studied theoretically and in simulation may not be all achievable for practical reasons:
\begin{itemize}
    \item[1.] The distance $z$ has to fit inside the instrument you are designing. In practice, we can assume that $z$ can only vary from a few centimeters at the minimum to a few meters at the very most. 
    \item[2.] The IAP is set by DM construction techniques. An IAP of $0.3$ mm and $1$ mm is currently or will soon be manufactarable for up to at least $64 \times 64$ actuators, but smaller or intermediary IAPs are currently not. 
    \item[3.] The wavelength is often determined by the science case (young self-emitting planets are bright in IR, but cooler planets require instruments in visible)
\end{itemize}

For these reasons, given that the best performance is achieved for Fresnel numbers between $10^{2}$ and $10^{3}$, for visible wavelengths it is preferable to use small IAP DMs. To get to the best performance for the WFIRST aperture and a 3-10 $\lambda_0/D_{ap}$ DH ($F_0= 5\times 10^{2}$) using large Xinetics DMs (IAP = 1 mm, D = $48*1$ mm), such as the ones currently used in the WFIRST setup, would require a $z$ propagation distance of 8 meters. The use of smaller DMs could reduce this distance to under a meter. However, other factors (e.g. DM surface quality) can also drive the choice of the DMs.

We also note that the current best Fresnel number ($F_0= 5.4\times10^2$) corresponds to the limit of what is currently achievable (BMC like DMs separated by 0.7m). For this reason, and because it is unlikely that coronagraphic instruments will be built at shorter-than-visible wavelengths, we do not think that smaller DMs than the ones currently on the market are necessary. However, a more detailed study on the position of the Fresnel optimal point for several DH sizes, numbers of actuators, BWs and aperture sizes is necessary.

%-------------------------------------------------------------------------------------------------
%-------------------------------------------------------------------------------------------------
%-------------------------------------------------------------------------------------------------
\section{Conclusion}
%-------------------------------------------------------------------------------------------------
%-------------------------------------------------------------------------------------------------
%-------------------------------------------------------------------------------------------------

In this paper, we used the technique introduced in ACAD-OSM I, and explored several tools to optimize this technique in the context of future missions. In the first part, we showed how the choice of coronagraph can change the performance, mainly in terms of throughput. We then introduced two different tools (size of the Lyot stop, number of interaction matrices) allowing us to finely tune the performance in contrast or in throughput at the expanse of the other. These parameters must be taken into consideration in the global optimization of a two deformable mirror + coronagraph system. 

The main result of this paper is the formalism we developed to explain the limitations in contrast and throughput of the two DM propagation methods, allowing us to predict and understand the obtained numerical simulation results. We described two opposing effects (vignetting by the second DM and Talbot effect) and explained how and in which cases these effects will impact the performance. This formalism is not specific to this method and can be used to understand all two deformable mirror techniques. We used these effects to explain the presence, in the numerical simulations of this paper, of an optimal DM setup, only dependent of the Fresnel number and not on the size of the DMs or their position alone. This point corresponds both to an optimum in contrast and in throughput. A significant consequence of this discovery is that the deformable mirror setup currently chosen for the WFIRST mission is probably optimal for neither contrast nor throughput performance. 

One major result of both ACAD-OSM I and the current paper is the continuity we observe between methods developed to correct for ``small'' phase and amplitude aberrations and the one presented here to correct for aperture discontinuities. Indeed, we show in ACAD-OSM I that this technique using an optimized linear technique obtained better performance than more complex techniques that try to solve the non-linear equations \citep{pueyo13}. Even more surprising, once the local minimum in contrast obtained with the ACAD-OSM method was found for heavily discontinuous apertures, we were able to verify theoretical relations between contrast and spectral bandwidth (ACAD-OSM I) made by \cite{shaklan06} in the context of ``small'' phase and amplitude aberrations. Finally, the formalism introduced in Appendix~\ref{sec:annex} can be applied to both ``small'' amplitude aberrations or discontinuities. \cite{beaulieu17} have started a similar study with ``small'' phase and amplitude aberrations and also show the influence of the DM setup on the performance. These two simultaneous approaches can converge and opens the possibility of an optimal instrument that can correct for both aberrations in the wavefront and discontinuities. 

Some improvements can be made to this technique. First, we need to understand precisely how the optimum in Fresnel number is impacted by parameters such as size of the DH, number of actuators, type of aperture, and spectral bandwidth. Second, we need to prove that this technique can be used with fixed mirror apodizing techniques \citep[PIAA, PAVC, ][]{Guyon14,fogarty17} to further enhance its capabilities in terms of throughput. 

Future laboratory tests on the High-Contrast Imager for Complex Aperture Telescopes \citep[HiCAT, ][]{ndiaye15} optical optical testbed at STScI will enable the experimental confirmation of the ACAD-OSM technique.

\acknowledgments{Johan Mazoyer is currently paid by NSF. This material is based upon work carried out under subcontract \#1496556 with the Jet Propulsion Laboratory funded by NASA and administered by the California Institute of Technology. The authors would like to thank John Krist (JPL), Pierre Baudoz (LESIA) and Sylvain Egron (STScI/ONERA/LAM) for valuable discussions. We are grateful to the referee for his very constructive inputs that have greatly improved the overall presentation and clarity of this paper.}

\appendix

\section{Appendix A: two DM propagation in the Talbot formalism}
\label{sec:annex}
Most of the tools used in this appendix (Talbot effect, frequency folding) have been already used 10 years ago in several papers \citep{shaklan06,giveon06,pueyo07}. However, they were usually used to determine surface quality limits of the optics and/or applied to DM optical design in the proposed Michelson interferometer configuration. In this appendix, we seek to find, in the specific case of two sequential DMs, the limitation of contrast, first in the case of very small strokes and then in the general case.

\subsection{Appendix A.1: Small strokes}
We assume that there are only amplitude errors in the aperture. For a given number $N$, number of cycles inside the aperture of diameter $D_{ap}$,  the field in the pupil plane (z = PP) writes as:
\begin{equation}
\label{eq:field_aperture}
E_{ap}(\lambda,N,z = PP) = A\left[1 + \epsilon[N] \cos\left( \dfrac{2 \pi N x}{D_{ap}}\right)\right]
\end{equation}
A is the amplitude of the field created by the discontinuities. $A\epsilon[N]$ is the $N^{th}$ coefficient of the decomposition of the discontinuities in Fourier series. It is therefore the amplitude  at $N \lambda_0/D_ap$ of the electrical field created by this aperture. Because only frequencies inside the DH are important here, we assume $N \le OWA$. We call: 
\begin{equation}
X =   \dfrac{2 \pi x}{D_{ap}}
\end{equation}
The first DM (in pupil plane) can introduce a phase $\phi_{dm1, N}$:
\begin{equation}
\phi_{DM1}(N) =  \dfrac{\lambda_0}{\lambda}\sigma_{DM1}\cos(NX)
\end{equation}
where $\lambda_0$ is the central wavelength and $\sigma_{DM1}$ is the strokes on the first DM, normalized by the central wavelength $\lambda_0$. Therefore the field introduced by the first DM in the pupil plane here is:
\begin{equation}
E_{DM1}(\lambda,N,z = PP) = \exp \left[ \dfrac{2 i \pi}{\lambda} \phi_{DM1} \right] = \exp \left[  \dfrac{2 i \pi \sigma_{DM1}\lambda_0  }{\lambda}   \cos(NX) \right]
\end{equation}
First, we assume that the field $A\epsilon[N]$ is small enough so that the strokes introduced by the DM to correct for it still verify $ \sigma_{DM1} \ll  1$. 

\begin{equation}
\label{eq:champDM1}
E_{DM1}(\lambda,N,z = PP) = 1+  \dfrac{2 i \pi \lambda_0  }{\lambda} \sigma_{DM1}  \cos(NX) + \mathcal{O}(\sigma_{DM1}^2)
\end{equation}
A second DM is located at a distance $z$. This DM also introduce only phase in its optical plane ($z = DM2$). We also assume $\sigma_{DM2} \ll  1$ here:
\begin{equation}
E_{DM2}(\lambda,N,z = DM2) = 1+  \dfrac{2 i \pi \lambda_0  }{\lambda}  \sigma_{DM2} \cos(NX) + \mathcal{O}(\sigma_{DM2}^2)
\end{equation}
The field introduced in the pupil plane by the second DM is \citep[p87-89 in][]{goodman05}:
\begin{equation}
E_{DM2}(\lambda,N,z = PP) = 1+  \dfrac{2 i \pi \lambda_0  }{\lambda}\sigma_{DM2}   \cos(NX)\exp\left( - \dfrac{i \pi \lambda z N^2}{D^2}\right)
\end{equation}
This equation shows that for certain distance $z_t(\lambda,N)$, the field created by the second DM reconstructs itself. This distance is called the Talbot distance and is defined as :  
\begin{equation}
\label{eq:talbot2}
z_{t}(\lambda,N)  = \dfrac{2D^2}{\lambda N^2}\,\,,
\end{equation}
We recall the Fresnel number at the central wavelength $F_0$: 
\begin{equation}
F_0 = \dfrac{D^2}{\lambda_0 z} 
\end{equation}
and we can write the electrical field $E_{DM2}$ as:
\begin{equation}
E_{DM2}(\lambda,N,z = PP) = 1+  \dfrac{2 i \pi \sigma_{DM2}\lambda_0  }{\lambda}   \cos(NX)\exp\left( - \dfrac{i \lambda \pi N^2}{ \lambda_0 F_0}\right) 
\end{equation}
We define the Talbot-limited range as the configuration in which $F_0\gg N^2$. This corresponds to the case where the corrected frequency $N$ is small compared to the Half-Talbot (``worst'') frequency $N_{t/2} = \sqrt{F_0}$ (Eq.~\ref{eq:freq_talbot}). In this regime, we find:

\begin{equation}
\label{eq:talbot_approximation}
 E_{DM2}(\lambda,N,z = PP) = 1+  \dfrac{2 i \pi\lambda_0  }{\lambda} \sigma_{DM2}  \cos(NX)\left( 1 - \dfrac{i\lambda \pi N^2}{ \lambda_0 F_0} + \dfrac{1}{2}\left(\dfrac{\pi \lambda N^2}{\lambda_0 F_0}\right)^2 \right)
\end{equation}
In \cite{mazoyer17SPIEfun}, we show that it is possible to develop analytically this calculation without doing this limited development. However, in this paper, we are assuming $F_0\gg N^2$. The on-axis part is removed by the coronagraph and we only analyze the term in $\cos(NX)$. The field in pupil plane (we always assume z = PP from now on) is now, separated in real and imaginary part: 

\begin{equation}
\label{eq:champDM2_withoutfrequencyfolder}
E_{Tot}(\lambda,N) = \left[A\epsilon[N] + \dfrac{2 \pi^2 N^2}{F_0} \sigma_{DM2} \,\,\,\,\,\, , \,\,\,\,\,\, \dfrac{2 \pi \lambda_0  }{\lambda} (\sigma_{DM1} +\sigma_{DM2}) + \dfrac{ \pi^3 N^4  }{F_0^2}  \dfrac{\lambda}{\lambda_0} \sigma_{DM2}\right]
\end{equation}
The second DM is correcting for the amplitude:
\begin{equation}
A\epsilon[N] + \dfrac{2 \pi^2 N^2}{F_0} \sigma_{DM2} = 0
\end{equation}
The dependence in $\lambda$ is the same for the correcting term and for aberration to correct, therefore all amplitude at every wavelengths is corrected when:
\begin{equation}
\label{eq:sigma_DM2}
\sigma_{DM2} = - \dfrac{A\epsilon[N] F_0}{2  \pi^2 N^2}
\end{equation}
This formula explains the increase of the strokes with $F_0$ observed in the in the Talbot limited range. We now need to correct the phase with the first DM. However, some of the terms we seek to correct do not have the same wavelength dependence as the correction term in $\sigma_{DM1}$. Indeed, $\sigma_{DM1}$ has to verify:
\begin{equation}
\label{eq:imagin_part}
 \dfrac{2 \pi \lambda_0  }{\lambda} (\sigma_{DM1} +\sigma_{DM2}) + \dfrac{ \pi^3 N^4  }{F_0^2}  \dfrac{\lambda}{\lambda_0} \sigma_{DM2} = 0
\end{equation}
The first DM has to correct for the phase introduced by the second DM (same chromatic dependence, possible at all wavelengths), but also for the phase introduced by the second order of the Talbot effect (different chromatic dependence). When we replace $\sigma_{DM2}$ by its value, Eq.~\ref{eq:imagin_part} reads:
\begin{equation}
\dfrac{2 \pi \lambda_0  }{\lambda} (\sigma_{DM1} +\sigma_{DM2}) - A\epsilon[N] \dfrac{ \pi N^2  }{2F_0}\dfrac{\lambda}{\lambda_0}   = 0
\end{equation}
$\sigma_{DM1}$ cannot correct for this field at every wavelengths. We only correct it at the central wavelength $\lambda_0$:
\begin{equation}
\label{eq:sigma_DM1}
\sigma_{DM1}  = A\epsilon[N] \left(\dfrac{ N^2}{4F_0} - \dfrac{F_0}{2  \pi^2 N^2} \right)
\end{equation}
This means that, in absolute value, the strokes on the first DM will be less important than on the second DM although close in the Talbot-limited range ($F_0\gg N^2$). We finally write the residual (non corrected) field by replacing $\sigma_{DM1}$ by its value in the field in Eq.~\ref{eq:champDM2_withoutfrequencyfolder}: 

\begin{equation}
\label{eq:Eres}
E_{res}(\lambda,N) =  A\epsilon[N] i \dfrac{ \pi N^2  }{2F_0} \left[ \dfrac{\lambda_0  }{\lambda} - \dfrac{\lambda}{\lambda_0} \right]
\end{equation}
The contrast is the residual light level $|E_{res}|^2$ divided by the light that is stripped away by the coronagraph ($|A|^2$):
\begin{equation}
C_1(\lambda,N) = \epsilon[N]^2 \dfrac{ \pi^2 N^4  }{4F_0^2} \left[ \dfrac{\lambda_0  }{\lambda} - \dfrac{\lambda}{\lambda_0} \right]^2
\end{equation}
We integrate over the bandwidth and obtain:
\begin{equation}
C_1(N) = \epsilon[N]^2 \dfrac{ \pi^2 N^4  }{4F_0^2} \left(\dfrac{1}{4R^2 - 1} + \dfrac{1}{12R^2} \right)
\end{equation}
where R = $\Delta\lambda/\lambda_0$ is the spectral resolution. We finally integrate over the DH. We do not know precisely the aperture distribution $\epsilon[N]$ at each frequency. However, we know that this spectral distribution is decreasing with N. Therefore, for every N inside the DH $\epsilon[N] > \epsilon[OWA]$, which gives us a minimum boundary for the contrast over the DH:
\begin{equation}
C_{DH,1} = \epsilon[OWA]^2 \dfrac{ \pi^2 (OWA^5 -IWA^5) }{\Delta N 20F_0^2} \left(\dfrac{1}{4R^2 - 1} + \dfrac{1}{12R^2} \right) 
\end{equation}
Where $\Delta N = OWA - IWA$Finally, because $OWA^5 \gg IWA^5$ and $R \gg 1$
\begin{equation}
\label{eq:contrast_ini}
C_{DH,1} \simeq \epsilon[OWA]^2 \dfrac{ \pi^2 OWA^5  }{ 60 F_0^2 \Delta N} \dfrac{1}{R^2} 
\end{equation}

{This formula for contrast assumes that the strokes are small ($\sigma_{DM1}\ll 1$, $\sigma_{DM2}\ll 1$, approximation in Eq.~\ref{eq:champDM1}) and/or that we have a finite number of actuators. For a finite number of actuators (a finite DH), \cite{giveon06,pueyo07} showed that  the frequency folding term eventually limits the contrast. We showed in Eq.~\ref{eq:sigma_DM2} and \ref{eq:sigma_DM1} that the strokes necessary to correct for the amplitude aberrations are linearly dependent on F:}
\begin{equation}
\label{eq:approxsigma}
\sigma_{DM1}\sim - \sigma_{DM2} \sim  \dfrac{A\epsilon[N] F_0}{N^2}
\end{equation}
{This means that for every given amplitude aberrations in the entrance pupil, for a large Fresnel number, the strokes will eventually grow large enough so that the frequency folding term is the main contrast limitation.  We study this effect in the next section.}

\subsection{Appendix A.2: Larger strokes and frequency folding case}
From Eq.~\ref{eq:champDM1}, we develop to the next term:
\begin{align*}
E_{DM1}(\lambda,N) &= 1+  \dfrac{2 i \pi \lambda_0  }{\lambda} \sigma_{DM1}  \cos(NX) + \dfrac{1}{2}\left(\dfrac{2 i \pi \lambda_0  }{\lambda} \sigma_{DM1}\right)^2  \cos^2(NX) + \mathcal{O}(\left( \dfrac{A\epsilon[N] F_0}{N^2}\right)^3)\\
E_{DM1}(\lambda,N) &= 1 +\dfrac{1}{4}\left(\dfrac{2 i \pi \lambda_0  }{\lambda} \sigma_{DM1}\right)^2+  \dfrac{2 i \pi \lambda_0  }{\lambda} \sigma_{DM1}  \cos(NX) + \dfrac{1}{4}\left(\dfrac{2 i \pi \lambda_0  }{\lambda} \sigma_{DM1}\right)^2  \cos(2NX)+ \mathcal{O}(\left( \dfrac{A\epsilon[N] F_0}{N^2}\right)^3)
\end{align*}
This new term has an impact on the speckle at half-frequency ($f = 1/(2N)$), which is the reason it is called frequency folding. However, if we write this last equation for the double frequency ($f = 2/N$), the frequency folding will introduce an amplitude term in 1/$\lambda^2$ in $\cos(NX)$. This term can be written as:
\begin{align*}
E_{ff,DM1}(\lambda,N) &= \dfrac{1}{4}\left(\dfrac{2 i \pi \lambda_0  }{\lambda} \sigma_{DM1}[f = 2/N]\right)^2 \cos(NX)\\
E_{ff,DM1}(\lambda,N) &= -(4\pi)^2\left(\dfrac{\lambda_0 }{\lambda}\right)^2\left(\dfrac{A\epsilon[N/2] F_0}{N^2}\right)^2 \cos(NX)
\end{align*}
using approximation for $\sigma_{DM1}[f = 2/N]$ in Eq.~\ref{eq:approxsigma}. We also have the same term for the second DM ($E_{ff,DM2} \simeq E_{ff,DM1}$). Once again we remove the on-axis term and only keeps the $\cos(NX)$ terms. The total field at the Nth frequency now becomes:
\begin{multline}
E_{Tot}(\lambda,N) = \Bigg[A\epsilon[N] + \dfrac{2 \pi^2 N^2}{F_0} \sigma_{DM2} + \left[1 - \dfrac{1}{2}\left(\dfrac{\pi \lambda N^2}{\lambda_0 F_0}\right)^2 \right](E_{ff,DM1} +E_{ff,DM2}) ,\\ 
\dfrac{2 \pi \lambda_0  }{\lambda} (\sigma_{DM1} +\sigma_{DM2})+  \dfrac{ \pi^3 N^4  }{F_0^2}  \dfrac{\lambda}{\lambda_0} \sigma_{DM2} + \dfrac{\lambda \pi N^2}{ \lambda_0 F_0}(E_{ff,DM1} +E_{ff,DM2})\Bigg]
\end{multline}
and we replace $E_{ff,DM1}$ and $E_{ff,DM2}$ by their values:
\begin{multline}
E_{Tot}(\lambda,N) = \Bigg[A\epsilon[N] + \dfrac{2 \pi^2 N^2}{F_0} \sigma_{DM2}  -  32 \pi^2 \left(\dfrac{\lambda_0 }{\lambda}\right)^2\left(  \dfrac{A\epsilon[N/2] F_0}{N^2}\right)^2 + (2\pi)^4 (A\epsilon[N/2])^2,\\ 
\dfrac{2 \pi \lambda_0  }{\lambda} (\sigma_{DM1} +\sigma_{DM2}) +  \dfrac{ \pi^3 N^4  }{F_0^2}  \dfrac{\lambda}{\lambda_0} \sigma_{DM2} +  32\pi^3 \dfrac{\lambda_0  }{\lambda} \dfrac{(A\epsilon[N/2])^2 F_0}{N^2}\Bigg] \label{eq:champDM2_withfrequencyfolder}
\end{multline}
{This equation has to be compared to Eq.~\ref{eq:champDM2_withoutfrequencyfolder}, where only the amplitude in the aperture $A$ was corrected for (real part of Eq.~\ref{eq:champDM2_withoutfrequencyfolder}). Here, for a correction with a finite number of actuators, the amplitude to be corrected for by the second DM is increasing as $F_0^2$. This shows that, whatever is the initial amplitude $A$ in the aperture, the amount of amplitude that effectively needs to be corrected by the 2 DM system depends on the DM setup and increases with $F_0$. We know from the ACAD-OSM I (Section 4), that more amplitude is impacting negatively both contrast and throughput. This is the reason why, in the Talbot limited range $F_0 \gg N^2$, the performance both in contrast and in throughput are decreasing with the Fresnel number. }

As in the previous case, we correct for the amplitude term (real part of $E_{Tot}$) with the second DM. The term in $A\epsilon[N]$ have been corrected in the last section. The term in $(A\epsilon)^2$ has the same chromatic dependence as the correction term in $\sigma_{DM2}$. It can be corrected entirely at all wavelengths with strokes negligible compare to Eq.~\ref{eq:sigma_DM1}. The real problem comes from the term in $(\lambda_0/\lambda)^2$. We can correct it at the central wavelength with:
\begin{equation}
\label{eq:strokeDM2_secondpart}
\sigma_{DM2,(\lambda_0/\lambda)^2} = - 16\dfrac{(A\epsilon[N/2])^2 F_0^3}{ N^6}
\end{equation}
and the residual field is:
\begin{equation}
E_{res_{ff},amp}= -  32 \pi^2\left(\dfrac{A\epsilon[N/2] F_0}{N^2}\right)^2 \left[ 1-\left( \dfrac{\lambda_0 }{\lambda}\right)^2 \right]
\end{equation}
Depending on the value of $A$ $\epsilon$, and $F_0/N^2$, this term is higher or not than the initial contrast $E_{res}$ in Eq.~\ref{eq:Eres}. However, whatever the initial amplitude, at large enough $F_0$ we have necessarily $|E_{res_{ff},amp}| \gg |E_{res}|$.\\

Now, just like previously, the strokes introduced on the second DM in Eq.~\ref{eq:strokeDM2_secondpart} must be corrected in the phase part (imaginary part of $E_{Tot}$) with the first DM. All the terms that have the $\lambda_0/\lambda$ dependence can be corrected at all wavelengths with strokes negligible compare to Eq.~\ref{eq:sigma_DM1}. The real problem comes from the correction of the $\lambda/\lambda_0$ term, that we can correct at the central wavelength with: 
\begin{equation}
\sigma_{DM1,\lambda/\lambda_0} = 8\pi^2 \dfrac{(A\epsilon[N/2])^2 F_0}{N^2}
\end{equation}
and the resulting field is 
\begin{equation}
E_{ff, pha}(N,\lambda)= 16 i \pi^3 (A\epsilon[N/2])^2 \dfrac{F_0 }{N^2} \left[ \dfrac{\lambda_0  }{\lambda} - \dfrac{\lambda}{\lambda_0} \right]
\end{equation}
Once again, depending on the value of $A$ $\epsilon$, and $F_0/N^2$, this term is higher or not than the initial contrast $E_{res}$ in Eq.~\ref{eq:Eres}. However, whatever is the initial amplitude, at large enough $F_0$ we have necessarily $|E_{res_{ff},pha}| \gg |E_{res}|$. Because $E_{res_{ff},amp}$ is only real and $E_{res_{ff},pha}$ is only imaginary, the final contrast is the sum of the contrasts created by these two fields independently. \\

For $E_{res_{ff},amp}$, the contrast is:
\begin{equation}
C_{2}(N,\lambda)=  \left(32 \pi^2\right)^2 \left(\dfrac{A\epsilon[N/2] F_0}{N^2}\right)^4 \left[ 1-\left( \dfrac{\lambda_0 }{\lambda}\right)^2 \right]^2
\end{equation}
We integrate over the bandwidth:
\begin{equation}
C_{2} (N)=  \left(32 A \pi^2\right)^2 \left(\dfrac{\epsilon[N/2] F_0}{N^2}\right)^4 \left[  \dfrac{1}{3 R^2} + \dfrac{1}{80 R^4}\right] 
\end{equation}
Once again, for every N inside the DH $\epsilon[N/2] > \epsilon[OWA]$, which gives us a minimum boundary for the contrast after integration over the DH:
\begin{equation}
C_{DH,2} =   \left(32 \pi^2\right)^2  (\epsilon[OWA] F_0)^4 \left( \dfrac{1}{3 R^2} + \dfrac{1}{80 R^4}\right) \left( \dfrac{1}{7IWA^7} - \dfrac{1}{7OWA^7}\right)
\end{equation}
Finally, because $OWA^7 \gg IWA^7$ and $R \gg 1$
\begin{equation}
\label{eq:Cres_apm}
C_{DH,2}  \simeq \dfrac{1024 \pi^4  }{21} A^2 (\epsilon[OWA] F_0)^4  \dfrac{1}{R^2}\dfrac{1}{IWA^7} 
\end{equation}
And for $E_{res_{ff},pha}$, the contrast reads:
\begin{equation}
C_{3}(N,\lambda)=  256\pi^6  (A\epsilon[N/2])^4 \left(\dfrac{F_0 }{N^2}\right)^2 \left[ \dfrac{\lambda_0  }{\lambda} - \dfrac{\lambda}{\lambda_0} \right]^2
\end{equation}
We also integrates and finally obtain:
\begin{equation}
\label{eq:Cres_pha}
C_{DH,3}  \simeq  \dfrac{256 \pi^6}{9}    A^2   \dfrac{F_0^2 }{R^2 }\dfrac{\epsilon[OWA]^4}{IWA^3}
\end{equation}

\subsection{Appendix A.3: Conclusion of Appendix A}

It is hard to compare $C_{DH,1}$, $C_{DH,2}$ and $C_{DH,3}$ in the general case, without specific values for $A$, $\epsilon$, $IWA$, $OWA$ and $F_0$. Also, in practice, the ACAD-OSM algorithm is not first correcting for the amplitude totally and then correcting for the frequency folding term but the two of them at the same time, depending which one is limiting contrast at a given iteration. 

However, in the Talbot-limited range ($F_0 \gg N^2$), the following points can be made:
\begin{itemize}
    \item For a finite number of actuators, the correction of the amplitude in the aperture only is not enough to increase contrast. This is the main reason why the ACAD-ROS method is barely improving contrast (see ACAD-OSM~I). Indeed, as soon as the DMs introduce strokes to correct for this aperture amplitude, they also introduce a frequency folding term (namely, the amplitude term created by the second order of the phase introduced on the first DM) that eventually limit the contrast. This term is increasing with $F_0$.
    \item Even for a small initial amplitude aberration in the aperture initially, the amplitude actually corrected for by the second DM is therefore increasing with $F_0$. (Eq.~\ref{eq:champDM2_withfrequencyfolder}). This increase of the amplitude to correct is the reason why the performance in throughput and in contrast decrease with $F_0$ in the Talbot-limited range ($F_0\gg N^2$).
    \item For Fresnel numbers or in the case of an infinite number of actuators, we are in the case of appendix A.1 and the contrast in the DH goes in $C_{DH,1} \sim OWA^5/F_0^2$.
    \item However, for larger Fresnel numbers, the frequency folding term is the main limitation of the contrast in the DH, which goes in goes in  $C_{DH,2} \sim F_0^2/IWA^3$ and then in  $C_{DH,3} \sim F_0^4/IWA^7$.
    \item the wavelength dependence in the Talbot regime is the same for all of these contrast term and is in $\sim 1/R^2$. This is the bandwidth dependence used in Eq.~2 of ACAD-OSM I.
\end{itemize}

\bibliographystyle{apj} 
\bibliography{acad_bib.bib}  

\end{document}